\documentclass[%
 reprint,
 amsmath,amssymb,
 $aps,
prb,
]{revtex4-1}

\usepackage{graphicx} 
\usepackage{bm}
\usepackage{txfonts}
\usepackage{mathrsfs}
\usepackage{amsmath}
\usepackage{xcolor}

\usepackage{subcaption} 

\def\b0{{\mathbf 0}}

\def\b0{{\mathbf 0}}

\def\tq{\tilde q}

\def\beq{\begin{equation}}
\def\eeq{\end{equation}}
\def\beqa\begin{eqnarray}
\def\eeqa{\end{eqnarray}}

\def\vecq{\vec{q}}

\def\trho{\tilde{\rho}} 
\def\tgam{\tilde{\gamma}}
\def\tu{\tilde{u}} 
\def\tw{\tilde{w}}

\begin{document}

\title{Stability of the Fulde-Ferrell-Larkin-Ovchinnikov states in anisotropic systems  and \\ critical behavior at thermal $m$-axial Lifshitz points}
\author{Piotr Zdybel}
\affiliation{Institute of Theoretical Physics, Faculty of Physics, University of Warsaw, Pasteura 5, 02-093 Warsaw, Poland}
\author{Mateusz Homenda}
\affiliation{Institute of Theoretical Physics, Faculty of Physics, University of Warsaw, Pasteura 5, 02-093 Warsaw, Poland}
\author{Andrzej Chlebicki}
\affiliation{Institute of Theoretical Physics, Faculty of Physics, University of Warsaw, Pasteura 5, 02-093 Warsaw, Poland}
\author{Pawel Jakubczyk }
\affiliation{Institute of Theoretical Physics, Faculty of Physics, University of Warsaw, Pasteura 5, 02-093 Warsaw, Poland} 
\date{\today}
\begin{abstract}
We revisit the question concerning stability of nonuniform superfluid states of the Fulde-Ferrell-Larkin-Ovchinnikov (FFLO) type to thermal and quantum fluctuations. Invoking the  properties of the putative phase diagram of two-component Fermi mixtures, on general grounds we argue, that for isotropic, continuum systems the phase diagram hosting a long-range-ordered FFLO-type phase envisaged by the mean-field theory cannot be stable to fluctuations at any temperature $T>0$ in any dimensionality $d<4$. In contrast, in layered unidirectional systems the lower critical dimension for the onset  of FFLO-type long-range order accompanied by a Lifshitz point at $T>0$ is $d=5/2$. In consequence, its occurrence is excluded in $d=2$, but not in $d=3$. We propose a relatively simple method, based on nonperturbative renormalization group to compute the critical exponents of the thermal $m$-axial Lifshitz point continuously varying $m$, spatial dimensionality $d$ and the number of order parameter components $N$. 
We point out the possibility of a robust, fine-tuning free  occurrence of a quantum Lifshitz point in the phase diagram of imbalanced Fermi mixtures. 
\end{abstract}

\pacs{}

\maketitle

\section{Introduction}
The rapid development of experimental techniques in cold-atom systems\cite{Bloch_2005, giorgini_theory_2008, Strinati_2018} gave rise to a revival of interest in unconventional fermionic superfluids exhibiting pairing at finite center of mass momentum, the so called Fulde-Ferrell-Larkin-Ovchinnikov states.\cite{fulde_superconductivity_1964, larkin_nonuniform_1965, Agterberg_2020} Such phases were since long discussed in solid-state physics contexts, but more recently were also predicted to arise in Fermi mixtures of ultracold atoms involving a population (and/or mass) imbalance between the two particle species forming the Cooper pairs. Due to the high level of controllability, this class of systems constitutes an interesting and promising platform for exploiting exotic superfluid phases including those of the FFLO type. 

 In a cold, two-component Fermi mixture increasing the concentration imbalance may serve to gradually mismatch the Fermi surfaces of the two atomic species, which suppresses pairing and ultimately drives the system to the normal, polarized metallic phase. The modulated FFLO-type superfluid was predicted to occur as an intermediate phase constituting an energetic compromise between the uniform, BCS-type superfluid and the polarized Fermi liquid. Extensive studies spread over years (see e.g. Refs. \onlinecite{He_2006,   Gubbels_2009, Baarsma_2010, radzihovsky_imbalanced_2010, Cai_2011, Baarsma_2013, Rosher_2015, Karmakar_2016, Kinnunen_2018, Pini_2021_2, Rammelmuller_2021}) addressed the energetic aspects of the problem, in particular the competition between the different candidate modulated ground states. The emergent consensus is that (at mean field level) a superfluid pair-density-wave (FFLO) phase is rather robustly stable, albeit in a relatively narrow region of the phase diagram. 

Somewhat surprisingly, fluctuation effects occurring in the FFLO phases were addressed much less comprehensively and rather coherently pointed towards instability of these long-range-ordered pair-density-wave states to thermal order-parameter fluctuations.\cite{Shimahara_1998, Samokhin_2010, Radzihovsky_2009, Radzihovsky_2011, Yin_2014, Jakubczyk_2017, Wang_2018} The mechanism destabilizing the FFLO states is  somewhat akin to that prohibiting the long-range order in  $XY$ or Heisenberg ferromagnets in dimensionality $d\leq 2$. However, in contrast to the conventional  magnets, the FFLO phases involve not only rotational (superfluid) symmetry breaking, but also breaking of translational symmetry, leading to a significantly softer Goldstone fluctuation spectrum in the (putative) ordered phases. In consequence, the possibilities of realizing such symmetry-breaking states at $T>0$ is severely restricted also in $d=3$.  

The previous studies of Refs. \onlinecite{Shimahara_1998, Samokhin_2010, Radzihovsky_2009, Radzihovsky_2011, Yin_2014, Jakubczyk_2017, Wang_2018} departed from a (putative) FFLO type state and investigated the low-energy fluctuations around it. In other words, they addressed the question of stability of the modulated phase without recourse to global features of the phase diagram. 
In contrast, our present approach implements a different logical line  invoking the properties of the mean-field (MF) phase diagram, which, by necessity involves the presence of both uniform and nonuniform phases leading to the emergence of  a thermal Lifshitz point where the normal (Fermi liquid like), the modulated (FFLO), and the uniform superfluid (BCS-like) phases all coexist. We point out that the stability conditions for the Lifshitz point at $T>0$ to fluctuations are significantly stronger as compared to those of the FFLO phase alone. For the isotropic systems we demonstrate the instability of the Lifshitz point to order-parameter fluctuations in $T>0$ at any dimensionality $d<4$, which should in consequence lead to a complete suppression of the pair-density wave phase at any $T>0$. 
Our conclusion is therefore stronger than those reached in the previous studies. Moreover, our reasoning does not require any detailed knowledge concerning the FFLO-type state and the associated excitation spectrum above it. We argue on the other hand that the FFLO-type states of the uniaxial type, such as coupled arrays of atomic tubes considered e.g. in Refs.~\onlinecite{Lutchyn_2011, Revelle_2016, Sundar_2020} are stable in dimensionality $d=3$ (but not in $d=2$) and present themselves as plausible candidates for hosting the FFLO phases. 

We also point out that the situation at $T=0$ is entirely different. 
We predict that the FFLO state may well be a stable ground state in a range of values of the imbalance parameters, squashed (at $T=0$) between the BCS-type superfluid and normal metallic phases.  The emergent structure of the phase diagram at $T\geq 0$ implies the generic existence of a point located at $T=0$, where the three phases meet (i.e. the quantum Lifshitz point). Notably, such a quantum Lifshitz point\cite{Zdybel_2020} should occur as a fluctuation-driven entity without any need for fine-tuning of the system parameters.      

The outline of this paper is as follows:  In Sec.~II we summarize the standard model to describe Fermi mixtures with population/mass imbalance (applicable in both the solid-state and cold-atom contexts) and give an overview of the features of the corresponding mean-field phase diagram. We in particular point out the generic occurrence of a thermal Lifshitz point located at a temperature $T_L>0$. We subsequently elucidate the structure of the effective action to describe the system in the vicinity of the Lifshitz point. In Sec.~III we discuss the stability of the Lifshitz point to order parameter fluctuations at Gaussian level depending on the system dimensionality $d$ and the anisotropy index $m$. In Sec.~IV we provide an estimate of the Lifshitz critical exponents from a truncation of functional renormalization group for general $d$, $m$ and number of order parameter components $N$. We in particular confirm the picture derived in Sec.~III and point out that the anomalous dimension associated with a class of spatial directions is negative. We summarize the paper in Sec.~V.                   

\section{Summary of the model and mean-field results}
The common point of departure for theoretical studies of imbalanced Fermi mixtures (applicable also to electronic systems) is provided by the grand canonical Hamiltonian 
\beq 
\mathcal{H} = \sum_{\vec{k}, \sigma} \xi_{\vec{k}, \sigma} c^\dagger_{\vec{k}, \sigma} c_{\vec{k}, \sigma} + \frac{g}{V}\sum_{\vec{k}, \vec{k}', \vec{q}} c^\dagger_{\vec{k}+\frac{\vec{q}}{2}, \uparrow} c^\dagger_{-\vec{k}+\frac{\vec{q}}{2}, \downarrow} c_{\vec{k}'+\frac{\vec{q}}{2}, \downarrow} c_{-\vec{k}'+\frac{\vec{q}}{2}, \uparrow} 
\label{Ham}
\eeq 
involving the kinetic energy term with species-dependent dispersion relation and chemical potential 
\beq 
\xi_{\vec{k}, \sigma} =\epsilon_{\vec{k}, \sigma}-\mu_\sigma 
\eeq 
 as well as  an attractive two-body interaction potential  driving $s$-waver pairing. For convenience the latter is taken in the form of a point-like interaction with $g<0$.  For the isotropic cold-atomic gases the dispersion reads $\epsilon_{\vec{k}, \sigma}=\frac{\vec{k}^2}{2 \mathcal{M}_\sigma}$, where the masses $\mathcal{M}_\sigma$ of the two fermionic species may in general be different. As we argue below, anisotropic kinetic terms allow for stabilizing the FFLO phases. This conclusion actually seems in line with experimental findings, since the most convincing evidence for FFLO-like features was reported for highly anisotropic situations both in the solid state\cite{Uji_2012, Uji_2013, Tsuchiya_2015, Koutroulakis_2016, Cho_2021} and ultracold gases contexts. We therefore do not restrict to any specific form of the dispersion. We nonetheless have in mind the setup, where $\tilde{m}$ out of the $d$ spatial directions are distinct from the remaining $d-\tilde{m}$. 
 As an experimentally relevant case one may, for example, invoke the following dispersion
 \beq 
 \epsilon_{\vec{k},\sigma}=    \sum_{i=1}^{\tilde{m}}  \frac{{k_i}^2}{2 \mathcal{M}_\sigma}  - 2 t_\perp \sum_{i=\tilde{m}+1}^d \cos k_i\;. 
 \label{Anizo_disp}
 \eeq      
In particular, for $d=2$ and $\tilde{m}=1$ this dispersion 
was implemented to describe coupled atomic tubes\cite{Lutchyn_2011} in the cold-atom context as well as specific organic superconductors.\cite{Mayaffre_2014, Piazza_2016} Note that the corresponding Fermi surfaces exhibit a considerable degree of nesting, favoring finite-momentum pairing. Clearly, for $\tilde{m}=d$ we recover from Eq.~(\ref{Anizo_disp}) the standard continuum gas, while $\tilde{m}=0$ yields the hipercubic lattice dispersion. One virtue of the parametrization is that both $d$ and $\tilde{m}$ ($\tilde{m}\leq d$) may formally be treated as real parameters, providing a way of continuously interpolating between different physically relevant cases (e.g. $\tilde{m}=0$ and $\tilde{m}=1$), see Sec.~III and IV.   
\begin{figure}[ht] 
\begin{center}
\includegraphics[width=9cm]{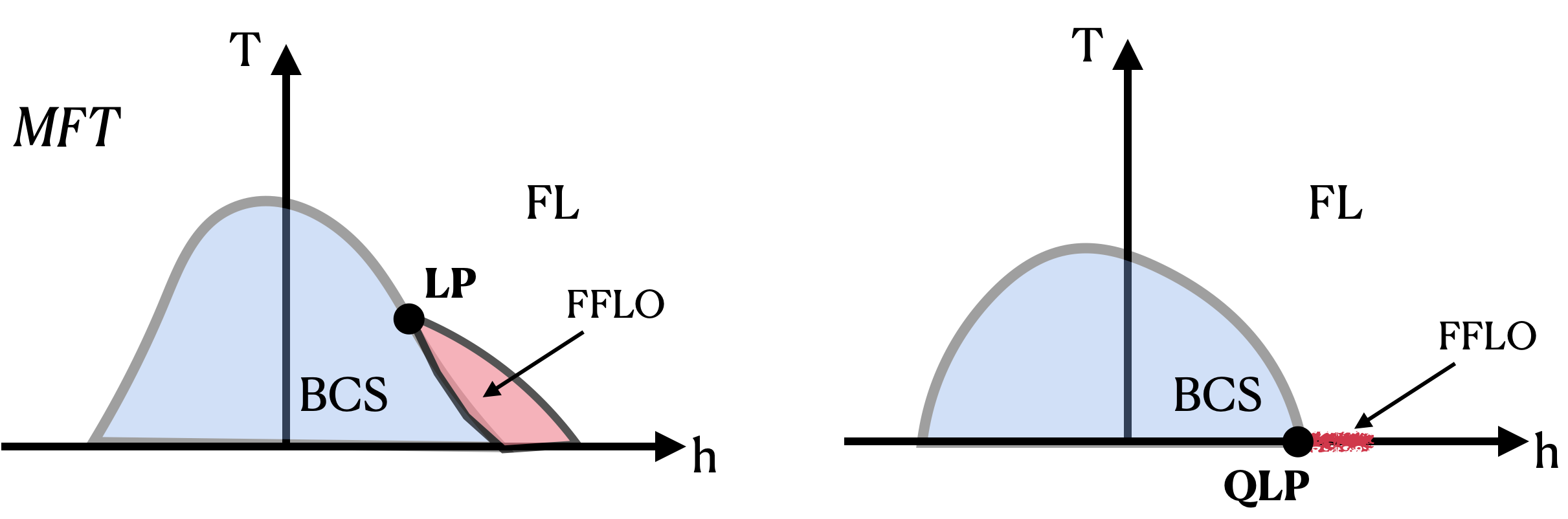}
\caption{(Color online) Left panel: schematic mean-field phase diagram of a system described by the Hamiltonian of Eq.~(\ref{Ham}). Increasing the imbalance parameter $h=(\mu_\uparrow -\mu_\downarrow)/2$ suppresses pairing. The pair density wave (FFLO)  phase is energetically favored in a (typically tiny) region between the uniform superfluid (BCS) and Fermi-liquid (FL) states. The Lifshitz point where these three phases coexist, is inevitably present at $T>0$ and constitutes the bottleneck for stability of the phase diagram with respect to fluctuations. Right panel: The anticipated renormalized phase diagram of an isotropic system, where the FFLO phase survives only at $T=0$ and a quantum Lifshitz point occurs (see the main text). }
\label{Phase_diag}
\end{center} 
\end{figure} 
\subsection{Pairing susceptibility}
We will now approach the pairing instability from the symmetric (Fermi-liquid) phase (compare Fig.~1) corresponding to sufficiently large imbalance parameter $h=(\mu_\uparrow - \mu_\downarrow)/2$ and/or temperature $T$. 
The effective (Landau-Ginzburg) order-parameter action for the model given by Eq.~(\ref{Ham}) can be constructed by a standard procedure described in literature (see e.g. Refs.~\onlinecite{Strack_2014, Piazza_2016, Zdybel_2018, Zdybel_2019}) analogous to the one developed long ago for magnetic transitions.\cite{Nagaosa_book} Up to terms quadratic in the pairing field $\phi$, the effective action reads
\beq
\mathcal{S}_{eff}^{(2)} = \sum_{\vec{q}, i\omega_n} \phi^*_{\vec{q}, i\omega_n} \left[ -1/g - \chi_0 (\vec{q}, i\omega_n) \right] \phi_{\vec{q}, i\omega_n}\;. 
\label{action}
\eeq 
Here $\phi_{\vec{q}, i\omega_n}$ is the (scalar) complex $s$-wave pairing field written in the momentum-frequency representation, while $\chi_0 (\vec{q}, i\omega_n)$ involves the particle-particle bubble and may be expressed as
\beq
\chi_0 (\vec{q}, i\omega_n) = T\int_{\vec{k}}\frac{1-f\left(\xi_{\vec{k}, \downarrow}\right)-f\left(\xi_{\vec{k}+\vec{q}, \uparrow}\right)}{\xi_{\vec{k}+\vec{q}, \uparrow}+\xi_{\vec{k}, \downarrow}-i\omega_n}\;,
\eeq
where  $f(X)=(e^{X/T}+1)^{-1}$ is the Fermi-Dirac distribution and $\int_{\vec{k}}=\int\frac{d\vec{k}}{(2\pi)^d}$.

 At mean-field level an instability towards superfluidity occurs once the Landau coefficient $a_2 =\left[ -1/g - \chi_0 (\vec{q}, 0) \right] $ in Eq.~(\ref{action}) becomes negative for some value of $\vec{q}$ (hereafter denoted as $\vec{Q}$). A nonzero ordering wavevector $\vec{Q}$ marks an FFLO-type instability. Note that $\chi_0 (\vec{q}, i\omega_n)$ involves no dependence on $g$. In consequence, once a set of parameters for which $\chi_0 (\vec{q}, 0)$ features a maximum at $\vec{q} = \vec{Q}\neq 0$  is identified, the (mean-field) transition between the normal and FFLO phases can be conveniently tuned by modifying $g$ alone.
\begin{figure}[ht] 
\begin{center}
\includegraphics[width=9cm]{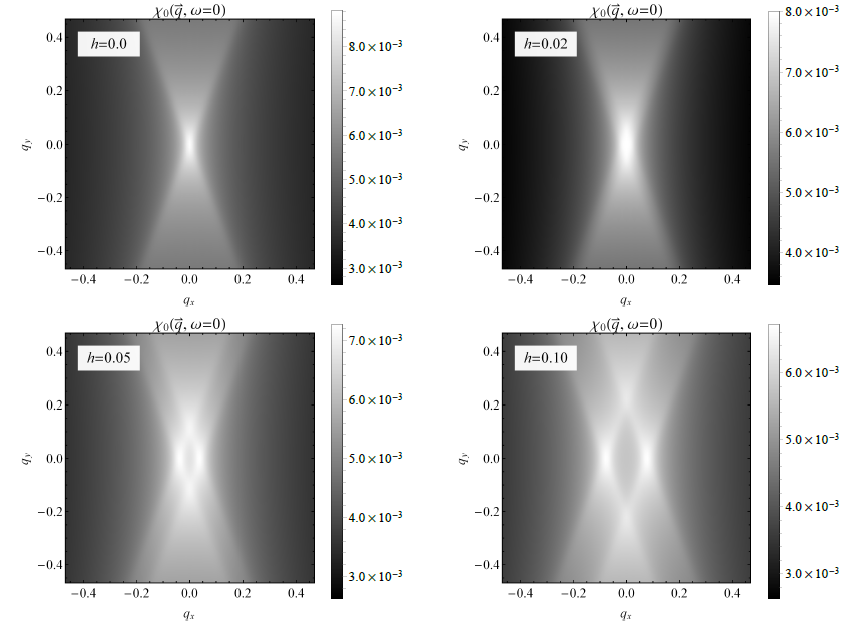}
\caption{Evolution of the pairing susceptibility $\chi_0 (\vec{q}, 0)$ upon varying $h$ for the dispersion given by Eq.~(\ref{Anizo_disp}) and $(d,\tilde{m})=(2,1)$. The sharp peak located for $h=0$ at $\vec{q}=(0,0)$ broadens and for $h=h_c\approx 0.03$ continuously splits. At $h=h_c$ and $q$ small, the  pairing field propagator is quadratic in momentum in the $q_y$ direction, but quartic along $q_x$. The degenerate maxima of $\chi_0 (\vec{q}, 0)$ are located at the $q_x$ axis and remain well separated from zero for $h>h_c$. For a projection on the $q_x$ axis, compare Fig.~\ref{chi_projection}. The plot parameters are $t_\perp=\frac{1}{2}$, $\mu_\uparrow+\mu_\downarrow=6.6$, $\mathcal{M}_\downarrow/\mathcal{M}_{\uparrow}=1.0$, $T=10^{-2}$.}
\label{chi}
\end{center} 
\end{figure} 
\begin{figure}[ht] 
\begin{center}
\includegraphics[width=9cm]{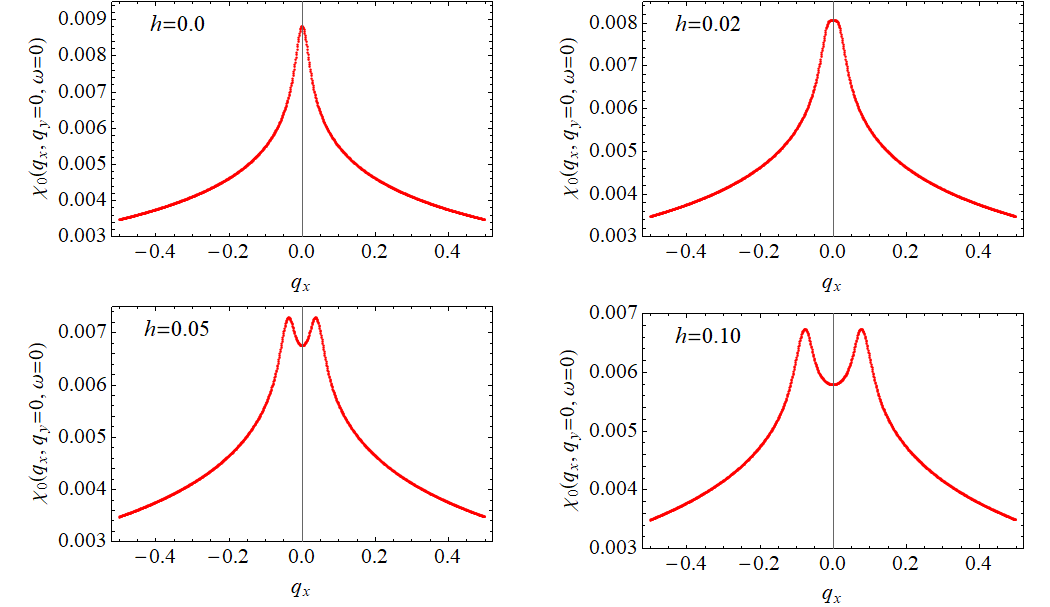}
\caption{Projection of the pairing susceptibility $\chi_0 (\vec{q}, 0)$ plotted in Fig.~\ref{chi} on $\vec{q}=(q_x,0)$.  }
\label{chi_projection}
\end{center} 
\end{figure} 
\begin{figure}[ht] 
\begin{center}
\includegraphics[width=9cm]{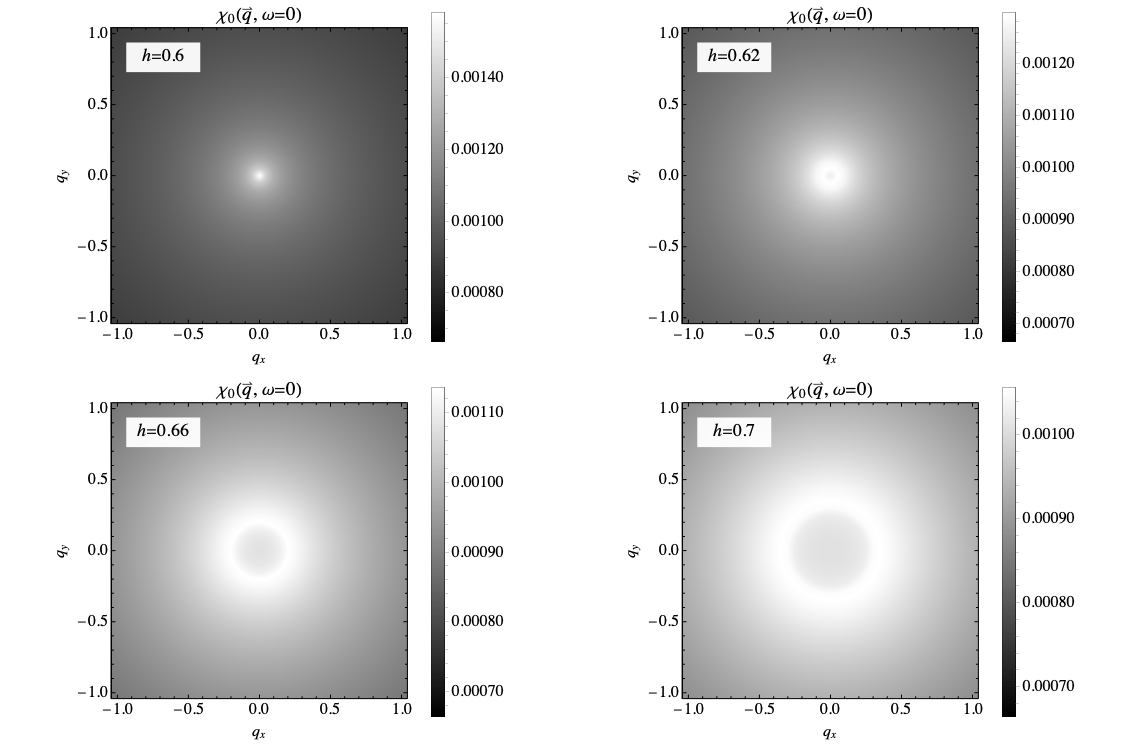}
\caption{Evolution of the pairing susceptibility $\chi_0 (\vec{q}, 0)$ upon varying $h$ for the dispersion given by Eq.~(\ref{Anizo_disp}) and $(d,\tilde{m})=(3,0)$ [i.e. for an isotropic continuum system of atomic particles]. The sharp peak located for $h$ sufficiently small at $\vec{q}=(0,0)$ broadens upon increasing $h$ and for $h=h_c\approx 0.62$ becomes degenerate on a two-dimensional sphere. At $h=h_c$ the pairing field propagator is quartic in momentum $\vec{q}$. We put $q_z=0$ in the plot.  A projection on the $q_x$ (or any other) axis, yields a picture qualitatively equivalent to the one presented in Fig.~\ref{chi_projection} for the anisotropic situation. The plot parameters are: $\mu_{\downarrow} +\mu_{\uparrow}=1$, $\mathcal{M}_{\downarrow}/\mathcal{M}_{\uparrow}=4.03$, $T=10^{-3}$ and were chosen to mimic the experimentally relevant $^{161}Dy$-$^{40}K$ mixture.\cite{Ravensbergen_2020}  }
\label{chi_iso}
\end{center} 
\end{figure} 
 The point $(\vec{Q}, 0)$ serves as a reference for the momentum/frequency expansion of the vertex functions in the Landau-Ginzburg action. In particular, the order parameter mass is given by 
\beq 
r = [-1/g -\chi_0(\vec{Q}, 0)]\;. 
\eeq
As already remarked, $\chi_0 (\vec{q}, 0)$ does not depend on $g$, which may therefore always be adjusted to obtain $r=0$. In Figs.~\ref{chi} and \ref{chi_iso} we plot $\chi_0 (\vec{q}, 0)$ for two situations corresponding to the uniaxial and isotropic cases. By varying $h$ or $T$ the position of the maximum (i.e. the ordering wavevector $\vec{Q}$) may be continuously shifted towards zero, where the superfluid phase becomes uniform [see Fig.~(\ref{Phase_diag})]. For clarity, in Fig.~\ref{chi_projection} we also expose the projection of $\chi_0 (\vec{q}, 0)$ plotted in Fig.~\ref{chi} on the $q_x$ axis.

In addition to the quadratic term given by Eq.~(\ref{action}) the effective action $\mathcal{S}_{eff}[\phi]$ involves order parameter self-interaction terms of order higher than two in the field $\phi$. 

Consider now the structure of the effective action approaching the Lifshitz point along the superfluid phase transition line (given by $r=0$) from above (compare Fig.~1). Upon crossing the Lifshitz point the ordering wavevector $\vec{Q}$ is continuously shifted away from zero and becomes degenerate. The degeneracy level is determined by the symmetry of the Fermi surface. For the isotropic case (see Fig.~\ref{chi_iso}) $\vec{Q}$ picks up any direction in the $(d-1)$-dimensional space. In the immediate vicinity of the Lifshitz point one may expand $\chi_0$ in momentum $\vec{q}$ around zero (retaining  terms up to quartic order). Right at the Lifshitz point the coefficients of at least some of the $q^2$ terms vanish and become negative below the Lifshitz point. Stability of the system is then retained due to the terms quartic in momentum.


\section{The Lifshitz point stability}
We now analyze the structure of the effective action for the order parameter $\phi$. 
From the expansion described above in Sec.~II A we obtain at the Lifshitz point an effective Landau-Ginzburg action which, when expressed in position representation, reads: 
\beq 
\mathcal{S}_{eff} =\int d^d x \left[U(|\phi|^2) +\frac{1}{2}Z_\perp \left(\nabla_\perp\phi\right)^2 +\frac{1}{2} Z_{||} \left(\Delta_{||} \phi\right)^2  \right]\;, 
\label{Seff}
\eeq 
where $U(|\phi|^2)$ denotes the effective potential, which may be expanded to yield a polynomial in $|\phi|^2$
\beq 
U(|\phi|^2) = r |\phi|^2 + u |\phi|^4 +\dots \,.
\eeq
The $\sim|\phi|^2$ term coefficient, resulting from Eq~(\ref{action})  vanishes at the entire phase transition line, including the Lifshitz point. The coefficients of the higher-order terms of $U(|\phi|^2)$ may be expressed by the fermionic loops evaluated at external momenta $\vec{q}=0$. The energy cost of creating order-parameter nonuniformities is governed by the laplacian terms in $m$ spatial directions and gradient terms in the remaining $d-m$ directions. Explicitly:  
\beq
 \left(\Delta_{||} \phi\right)^2 = \sum_{i=1}^2\sum_{\alpha,\,\beta =1}^m  \frac{\partial^2 \phi_i}{\partial x_\alpha \partial x_\beta}\frac{\partial^2 \phi_i}{\partial x_\alpha\partial x_\beta}
\eeq 
and
 \beq
\left(\nabla_\perp\phi\right)^2 = \sum_{i=1}^2\sum_{\alpha =m+1}^d \frac{\partial \phi_i}{\partial x_\alpha}\frac{\partial \phi_i}{\partial x_\alpha}\;,
\eeq
where the $i$ summation runs over the two  components of the pairing field $\phi$. For symmetry reasons the number $m$ of 'soft' directions in the action of Eq.~(\ref{Seff}) must either coincide with the anisotropy index $\tilde{m}$ of Eq.~\ref{Anizo_disp} or be equal $d-\tilde{m}$. 
Since, for the time being, we are interested in the thermal phase transition, we dropped the contributions from quantum fluctuations in Eq.~(\ref{Seff}).
In the above form $\mathcal{S}_{eff}$ accounts for a generic anisotropic situation, where the dispersion is quartic in $m$ ($m\leq d$) spatial directions and quadratic in the remaining ones (compare Sec.~II A). The isotropic case  corresponds to $m=d$. The above effective action describes the $m$-axial Lifshitz point, analogous to those studied previously in the contexts of anisotropic magnets 
\cite{Grest_1978, Selke_1988, Diehl_2002, Butera_2008} and liquid crystals.\cite{Chaikin_book, Singh_2000} One may now adopt the standard Gaussian level arguments \cite{Goldenfeld_book} to argue for the instability of the Lifshitz point with respect to order parameter fluctuations at sufficiently low dimensionality. Considering that the (putative) homogeneous ordered phase supports a massless transverse mode, from the structure of the effective action of Eq.~(\ref{Seff}) it follows that (in Fourier space) the (transverse) 2-point correlation function in the immediate vicinity of the Lifshitz point reads:
\beq 
G(\vec{q})=G(\vec{q}_\perp, \vec{q}_{||})=\frac{1}{Z_\perp \vec{q}_\perp^2 +Z_{||}(\vec{q}_{||}^2)^2}\;,
\eeq
while in real space 
\beq
G(\vec{x}_\perp, \vec{x}_{||}) \sim \int d^d q G(\vec{q})e^{i(\vec{q}_\perp \vec{x}_\perp + \vec{q}_{||} \vec{x}_{||})}=\int d^d q\frac{e^{i(\vec{q}_\perp \vec{x}_\perp + \vec{q}_{||} \vec{x}_{||})}}{Z_\perp \vec{q}_\perp^2 +Z_{||}(\vec{q}_{||}^2)^2}\,.
\eeq
By substituting $(\vec{q}_{||}^2)=\tilde{q}_{||}$, expanding the exponential occurring in the numerator, passing to spherical coordinates in each of the two subspaces and integrating over the angular coordinates, one arrives at the following expression: 
\beq 
\mathcal{C}\int_0^{\Lambda_\perp} dq_\perp \int_0^{\Lambda_{||}^2}d\tilde{q}_{||} \frac{\tilde{q}_{||}^{\frac{m}{2}-1} q_\perp^{d-m-1}}{Z_\perp \vec{q}_\perp^2 + Z_{||}\tilde{q}_{||}^2}
\eeq
with $\mathcal{C}$ constant and $\Lambda_\perp$, $\Lambda_{||}$ being microscopic (momentum) cutoffs. Transformation to polar coordinates: $\sqrt{Z_{||}}\tilde{q}_{||} = r\cos\phi$, $\sqrt{Z_\perp}q_\perp = r \sin{\phi}$ leads to an integral of the form 
\beq
G(\vec{x}_\perp, \vec{x}_{||})\sim\int_0^\Lambda dr r^{d-\frac{m}{2}-3}\;,
\eeq
divergent for $d\leq 2+\frac{m}{2}$. This implies instability of the Lifshitz point with respect to order parameter fluctuations for $d$ below $2+\frac{m}{2}$. The above treatment is analogous to a Gaussian level demonstration of the Mermin-Wagner theorem in the standard isotropic situations.\cite{Goldenfeld_book} 
The obtained condition coincides with those previously recognized for Lifshitz points for magnets and liquid crystals and may be cast in the form: 
\beq
d_L = 2+\frac{m}{2} \;,
\label{dL}
\eeq  
where $d_L$ is the lower critical dimension for occurrence of an $m$-axial Lifshitz point where the normal, homogeneous superfluid and FFLO phases would coexist. Note that for $m=0$ we recover $d_L=2$ in line with the Mermin-Wagner theorem,  while for the isotropic case ($m=d$) Eq.~(\ref{dL}) leads to $d_L=4$, which in dimensionality $d=3$ and $d=2$ prohibits the occurrence of the isotropic Lifshitz point, and in consequence also the FFLO phase squashed between the BCS-like and Fermi liquid phases according to the mean-field predictions. 

The expression of Eq.~(\ref{dL}) is not new and was first derived in the context of magnetic systems long ago by Grest and Sak\cite{Grest_1978} within the $2+\epsilon$ expansion of the nonlinear sigma model. It is expected to be valid for situations characterized by the number of order parameter components $N\geq 2$. Analysis of the Ginzburg criterion for the Lifshitz point\cite{Diehl_2002} indicates a similar effect on the upper critical dimension $d_u$, such that $d_u=4+\frac{m}{2}$.

As concerns the stability of the FFLO states, we emphasize the difference between the above arguments and those presented in earlier literature. While the previous studies addressed stability of different putative ground states of the FFLO type to Goldstone fluctuations, the present analysis invokes the envisaged presence of the thermal Lifshitz point in the phase diagram and inspects its stability to critical order parameter fluctuations. This yields in the isotropic situation a condition by far more restrictive. The absence of a thermodynamically stable long-range ordered FFLO phase certainly does not contradict the presence of regions of the phase diagram exhibiting enhanced FFLO pairing fluctuations (see Ref.~\onlinecite{Pini_2021} for a recent discussion) which may well be detected in experiments on various systems. On the other hand, our argument offers an explanation of why convincing experimental evidence for FFLO states was reported only for strongly anisotropic situations and demonstrates that the occurrence of a true long-range ordered FFLO thermodynamic phase in isotropic three-dimensional systems is in fact completely excluded. Also note that the FFLO phase may well remain stable at $T=0$ (compare Fig.~\ref{Phase_diag}) which implies the presence of a quantum Lifshitz point in addition to the FFLO quantum critical point\cite{Piazza_2016, Pimenov_2018} in the phase diagram.  

We also make the  observation that the condition for stability of the Lifshitz point is significantly weaker in the anisotropic case ($m<d$). For the most relevant uniaxial case $m=1$ one obtains $d_L=5/2$, which indicates a stable Lifshitz point in $d=3$, but not $d=2$. In this situation the stability restrictions for the FFLO phase with respect to Goldstone fluctuations (the spectrum of which depends on the details of the ground state)  are in fact more demanding,\cite{Radzihovsky_2011} implying in $d=3$ that the FFLO state may presumably appear as a quasi-long-range ordered phase.   

In the next section we provide further evidence, based on nonperturbative RG, which supports the above picture, and, in addition, yields estimates of the critical exponents at the Lifshitz point.  

\section{Lifshitz critical behavior} 
In this section we extend and deepen the above argumentation by analyzing the critical behavior at thermal $m$-axial Lifshitz points in $d$ dimensions beyond the Gaussian level. We generalize out  calculation by allowing for an arbitrary number of order parameter components $N$. It is worth emphasizing that a description of this class of critical phenomena by the tools of perturbative renormalization group (RG) is, as compared to the standard $O(N)$-symmetric case, way more demanding,\cite{Shpot_2001, Shpot_2005, Shpot_2008, Burgsmuller_2010, Shpot_2012}  and has up to now been carried out only up to two loop level. In the present chapter we implement a nonperturbative RG approach based on a truncation of the Wetterich equation. We begin by discussing the leading order of the derivative expansion, the so-called local potential approximation (LPA), which allows us to build a mapping between the Lifshitz and standard $O(N)$-symmetric critical behaviors. The discussed connection is exact at the LPA  level, but becomes violated after accounting for the anomalous dimensions, i.e. going to higher orders in the derivative expansion. This is here achieved with the simplest truncation in the spirit of the so-called LPA'. 
 The Lifshitz critical behavior was previously addressed with slightly different nonperturbative RG truncations in Ref.~\onlinecite{Essafi_2012} for the unaxial Heisenberg case in three dimensions $(d,N,m)=(3,3,1)$ and in Refs.~\onlinecite{Zappala_2017, Zappala_2018} for the isotropic case in $d\geq 4$. See also Ref.~\onlinecite{Defenu_2021} for an interesting connection between the isotropic Lifshitz point in $d=4$ and the Kosterlitz-Thouless transition. Our major current aim is to strengthen  the argumentation leading to the conclusion of Sec.~III and  build up a connection to the standard isotropic $O(N)$-symmetric models. We shall therefore consider $m$ varying continuously between zero and $d$. As a byproduct we provide estimates of the critical indices of the Lifshitz point characterized by general $(d,N,m)$ as obtained within our present relatively simple approach.   
\subsection{Wetterich equation and the derivative expansion} 
The Wetterich approach\cite{Wetterich_1993} constitutes a nonperturbative implementation of Wilsonian renormalization group. It relies on the exact RG flow equation
\begin{align}
\partial_k\Gamma_k[\phi] = \frac{1}{2}\textrm{Tr}\left\{\partial_k R_k \left[\Gamma_k^{(2)}[\phi] +R_k\right]^{-1}\right\}\;,
\label{Wetterich}
\end{align}
which evolves the effective average action $\Gamma_k[\phi]$ between the microscopic action $\mathcal{S}[\phi]$ and the thermodynamic free energy $\mathcal{F}[\phi]$ upon varying the infrared (momentum) cutoff parameter $k$ from the microscopic scale $k=\Lambda$ towards zero. Specifically: $\Gamma_{k\to\Lambda}[\phi]\to \mathcal{S}[\phi]$ and $\Gamma_{k\to 0}[\phi]\to \mathcal{F}[\phi]$. The infrared cutoff is implemented by adding a momentum-dependent function $R_k=R_k(\vec{q})$ to the inverse propagator, thus effectively damping  modes with momentum $q<k$, leaving the modes with $q>k$ untouched. The trace in Eq.~(\ref{Wetterich}) encompasses in the present context summation over momentum as well as components of the order-parameter field $\phi$, while $\Gamma_k^{(2)}[\phi] $ denotes the second functional derivative of $\Gamma_k[\phi]$. The framework resting upon Eq.~(\ref{Wetterich}) was over the last years fruitfully applied in a broad range of contexts (for reviews see for example Refs.~\onlinecite{Berges_2002, Pawlowski_2007, Kopietz_book, RG_book, Metzner_2012, Dupuis_2021}). 

One successful approximation scheme to integrate the Wetterich equation is recognized as the derivative expansion (DE). It amounts to classifying the symmetry-allowed terms occurring in $\Gamma_k[\phi]$  according to the number of field derivatives and truncating terms of order higher than a given value. This projects the functional differential equation Eq.~(\ref{Wetterich}) onto a finite, numerically manageable set of partial (integro-)differential flow equations. Only very recently was this framework systematically applied\cite{Polsi_2020}  to the case of $O(N)$-symmetric models at order $\partial^4$ (and  $\partial^6$ for the Ising universality class\cite{Balog_2019}) in dimensionality $d=3$. These computations led to estimates of the critical exponents of accuracy comparable to those delivered by the best Monte Carlo simulations and perturbation theory calculations. It will become clear that the case of the Lifshitz point constitutes a significantly more demanding challenge for the treatment based on the Wetterich approach. The reason for this is at least two-fold: (i) terms quartic in momentum appear in the inverse propagator even at the bare level and are crucial for capturing the relevant physics; (ii) the anisotropic nature of the problem complicates the loop integrals. Despite these, as we demonstrate below,  the Wetterich approach captures a substantial amount of physics and delivers estimates of the critical exponents even at the lowest orders of the DE.     
\subsection{Local potential approximation}
We now consider the leading (zeroth order) truncation of the derivative expansion, where the effective potential is a flowing (scale dependent) function, but the momentum dependencies in the propagator are not renormalized. This is known commonly as the local potential approximation (LPA) and, for the present problem, amounts to parametrizing $\Gamma_k[\phi]$ via the following form    
\beq 
 \Gamma_k[\phi]=\int d^d x \left[U_k(\rho) +\frac{1}{2}Z_\perp \left(\nabla_\perp\phi\right)^2 +\frac{1}{2} Z_{||} \left(\Delta_{||} \phi\right)^2  \right]\;, 
\label{LPA}
\eeq 
where we introduced $\rho =\frac{1}{2}|\phi|^2$. At this approximation level the gradient coefficients $Z_\perp$ and $ Z_{||}$ are scale independent (in consequence the anomalous dimensions are neglected), while there is no  preimposed parametrization of the flowing effective potential $U_k(\rho)$. Crucially, a term $\sim \left(\nabla_{||}\phi\right)^2$ is absent in Eq.~(\ref{LPA}). In a higher order calculation involving the flow of momentum dependencies of the propagator,  this term is present and should scale to zero at the Lifshitz point only for vanishing $k$.

   By plugging Eq.~(\ref{LPA}) into Eq.~(\ref{Wetterich}) we obtain a closed flow equation for $U_k(\rho)$ of the form     
\beq
\partial_k U_k(\rho) = \frac{1}{2}\int_q \partial_k R_k (\vec{q})\left[G_\sigma(\vec{q},\rho, m) + (N-1)G_\pi(\vec{q},\rho, m)\right]\;,
\label{LPA_eq}
\eeq
where 
\begin{align}
G_\sigma^{-1}(\vec{q},\rho, m)&= Z_{||}(\vec{q}_{||}^2)^2 + Z_\perp \vec{q}_\perp^2 + U_k'(\rho)+2\rho U_k''(\rho)+R_k(\vec{q}) \nonumber \\
G_\pi^{-1} (\vec{q},\rho, m)&= Z_{||}(\vec{q}_{||}^2)^2 + Z_\perp \vec{q}_\perp^2 + U_k'(\rho)+R_k(\vec{q}) 
\label{Gs}
\end{align}
are the regularized inverse propagators for the longitudinal ($\sigma$) and transverse ($\pi$) modes. The integral $\int_q = \int\frac{d^m q_{||}}{(2\pi)^m}\int \frac{d^{d-m} q_{\perp}}{(2\pi)^{d-m}}$ in Eq.~(\ref{LPA_eq}) encompasses the two subspaces characterized by distinct behavior of the dispersion. For $m=0$ we recover the standard LPA equation well studied for the $O(N)$-symmetric models, while for $m=d$ the $\vec{q}_{\perp}$-space is 0-dimensional which corresponds to the isotropic Lifshitz point. 

We now implement the following rescaling 
\begin{align}
\vec{q}_\perp = k \tilde{\vec{q}}_\perp\;,\; \vecq_{||}=(Z_\perp/Z_{\parallel})^{1/4} k^{1/2}\tilde{\vecq}_{||} \nonumber \\
\rho = Z_\perp^{\frac{m}{4}-1} Z_{\parallel}^{-\frac{m}{4}}k^{d-\frac{m}{2}-2}\tilde{\rho}\;,\; U_k(\tilde{\rho})=Z_\perp^{\frac{m}{4}} Z_{\parallel}^{-\frac{m}{4}}k^{d-\frac{m}{2}}\tilde{u}_k(\tilde{\rho})
\end{align} 
and consider the cutoff of the form 
\beq
R_k(\vec{q})=Z_{\perp}k^2 r\left(\tilde{\vec{q}}_\perp^2+(\tilde{\vecq}_{||}^2)^2\right)\;.
\eeq
This allows us to cast the LPA flow equation in a  scale invariant form:
\begin{align}
&\partial_t \tu_k = - \left(d-\frac{m}{2}\right)\tilde{u}_k - \left(2+\frac{m}{2}-d\right)\trho \tu_k'+\nonumber \\ 
&\frac{1}{2}\int_{\tq}\left[\frac{1}{y+\tu_k'+2\trho\tu_k''+r(y)} +\frac{N-1}{y+\tu_k' +r(y)} \right]\left[2r(y)-2yr'(y)\right]\;,
\end{align}
where we introduced $y=\tq_\perp^2 +\tq_{||}^4$ and $t=\log(k/\Lambda)$. In each of the two subspaces corresponding to $\tq_\perp$ and $\tq_{||}$ we now pass to the (hyper)spherical coordinates and perform the angular integrations. Subsequently the change of variables $\tq_\perp =\zeta \cos\theta$, $\tq_{||}^2=\zeta\sin\theta$ (with $\zeta=\sqrt{y}$) and integration over $\theta$ leads to the following form of the flow equation: 
\begin{align}
\partial_t \tu_k =& - \left(d-\frac{m}{2}\right)\tilde{u}_k- \left(2+\frac{m}{2}-d\right)\trho \tu_k'+\nonumber \\
&\mathcal{V}_{d,m}\int_0^\infty dy y^{\frac{d}{2}-\frac{m}{4}-1}\times\nonumber \\ 
&\left[2r(y)-2yr'(y)\right] \left[\frac{1}{y+\tu_k'+2\trho\tu_k''+r(y)} +\frac{N-1}{y+\tu_k' +r(y)} \right]\;, 
\label{LPA_resc}
\end{align}
with
\beq
\mathcal{V}_{d,m}=\frac{\mathcal{S}^{d-m-1}\mathcal{S}^{m-1}}{16(2\pi)^d}\mathcal{B}\left(\frac{d-m}{2},\frac{m}{4}\right)\;,
\eeq 
where in turn $\mathcal{S}^{n-1}=\frac{2\pi^{n/2}}{\Gamma(n/2)}$ is the surface area of the $(n-1)$-dimensional unit sphere, and $\mathcal{B}(x,y)$ denotes the Euler beta function.  

We now observe that by substituting $(d-\frac{m}{2})\rightarrow d_{eff}$ and $\mathcal{V}_{d,m}\rightarrow v_d= [2^{d+1}\pi^{d/2}\Gamma (d/2)]^{-1} $ in Eq.~(\ref{LPA_resc}) we recover the LPA equation for the standard $O(N)$-symmetric case in dimensionality $d_{eff}$. It follows that, at the LPA level of approximation, the RG equation for the $m$-axial Lifshitz point in $d$ dimensions differs from the corresponding flow equation for the $O(N)$ model in dimensionality $d_{eff}=d+\frac{m}{2}$ exclusively by the $m$-dependent constant multiplying the integral. The quantity $\mathcal{V}_{d,m}$ (and $v_d$ alike) is however redundant as it can be absorbed by the transformation 
\beq
u_k =   \mathcal{V}_{d,m} w_k \;,\;\;\;\; \tilde{\rho}=\mathcal{V}_{d,m}\tilde{\gamma}\;,
\eeq
which casts the LPA equation in the form 
\begin{align}
\partial_t \tw_k =& - \left(d-\frac{m}{2}\right)\tw_k- \left(2+\frac{m}{2}-d\right)\tgam \tw_k'+\nonumber \\
&\int_0^\infty dy y^{\frac{d}{2}-\frac{m}{4}}\times\nonumber \\ 
&\left[2r(y)-2yr'(y)\right] \left[\frac{1}{y+\tw_k'+2\tgam\tw_k''+r(y)} +\frac{N-1}{y+\tw_k' +r(y)} \right]\;, 
\label{LPA_resc_2}
\end{align}
where $\tw_k=\tw_k(\tgam)$ and prime now denotes differentiation with respect to $\tgam$. It follows that the critical behavior at the $m$-axial Lifshitz point is fully equivalent to the one at the $O(N)$-symmetric critical point at dimensionality reduced by $\frac{m}{2}$. This fact was previously recognized at the mean-field and Gaussian level as well as in the limit  $1/N\to 0$.\cite{Diehl_2002, Shpot_2008, Burgsmuller_2010, Shpot_2012, Jakubczyk_2018, Lebek_2020, Lebek_2021} The above reasoning indicates that the correspondence remains valid within the LPA approximation, and in fact should remain correct as long as the anomalous dimensions are neglected. 
In particular the lower (as well as the upper) critical dimension describing these two situations are then shifted by $m/2$ in agreement with the Gaussian argument presented in Sec.~III. In view of the above, the critical exponents for the Lifshitz point may be extracted using the routines previously developed for the $O(N)$ models. Here we focus on the $\nu_\perp$ exponent describing the decay of the correlation function $G(\vec{x}_\perp, \vec{x}_{||}=0)$. The analogous exponent $\nu_{||}$ controlling $G(\vec{x}_\perp=0, \vec{x}_{||})$ is related to $\nu_\perp$ via the scaling relation\cite{Diehl_2002} 
\beq
\nu_{||}=\frac{2-\eta_\perp}{4-\eta_{||}}\nu_\perp\;.
\eeq 
In the absence of anomalous dimensions we find $\nu_{||} =\frac{1}{2}\nu_\perp$. Other critical exponents are then also recovered via scaling relations.\cite{Diehl_2002}

    The quantity $\nu_\perp^{-1}$ may  be identified as the leading eigenvalue of the RG transformation of Eq.~(\ref{LPA_resc}) [or Eq.~(\ref{LPA_resc_2})] linearized around the fixed point. Technically, we first discretize Eq.~(\ref{LPA_resc}) on the $\trho$-grid (typically involving $\approx 60$ points) and solve for the fixed point $u^*(\trho)$. The RG equation Eq.~(\ref{LPA_resc}) is then linearized around $u^*(\trho)$ and its diagonalization yields the spectrum, which contains a single positive eigenvalue $\lambda_\nu$, which we identify with $\nu_\perp^{-1}$. For details on the numerical procedure, see Ref.~\onlinecite{Chlebicki_2021}. For $m\to 0$ we obviously recover the value pertinent to the standard $O(N)$ model. In the practical calculation we consider two families of cutoff functions: 
 \begin{align}
 r(y)= (1-y)\theta (1-y)\;\;\;\; &\textrm{(Litim cutoff)}  \\ 
 r(y)= \alpha \frac{y}{e^y-1} \;\;\;\; &\textrm{(Wetterich cutoff)}\;, 
 \label{cutoffs}
 \end{align}
 the latter one involving a variable parameter $\alpha$. The obtained value of $\nu_\perp^{-1}$ carries a weak dependence on $\alpha$. In accord with the principle of minimal sensitivity\cite{Canet_2003_2, Balog_2020} (PMS) one chooses $\alpha$ so that $\nu_\perp^{-1}$ is locally stationary with respect to variation of $\alpha$. Our results for $\nu_\perp$ depending on $N$, $d$ and $m$ are presented in Figs.~5-7 and compared with those obtained within the $\epsilon=4+\frac{m}{2}-d$ expansion and $\frac{1}{N}$ expansion. The differences between the values obtained using the different cutoff functions are relatively small and here we present the results obtained using the Litim cutoff. 
\begin{figure}[ht] 
\begin{center}
\includegraphics[width=9cm]{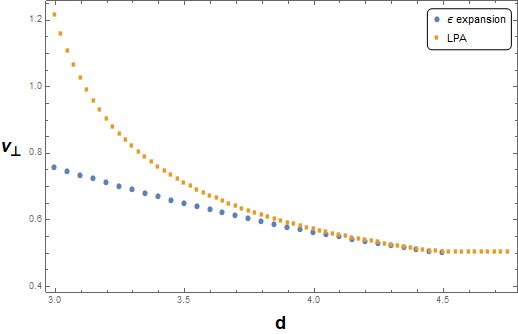}
\caption{The correlation length exponent $\nu_\perp$ for the uniaxial ($m=1$) Lifshitz point for $N=2$ plotted as a function of dimensionality $d$. The results obtained within the LPA approximation are superimposed with those resulting from the $\epsilon=4\frac{1}{2}-d$ expansion in Ref.~(\onlinecite{Shpot_2001}) up to order $\epsilon^2$. The two sets of points coincide in the vicinity of the upper critical dimension $d_u=4\frac{1}{2}$, above which we recover the mean-field result $\nu_\perp=\frac{1}{2}$. An increase of $\nu_\perp$ upon lowering $d$, indicating the expected divergence at the lower critical dimension $d_L=2\frac{1}{2}$ is clearly visible in the LPA data. }
\label{LPA_vs_eps}
\end{center} 
\end{figure} 
\begin{figure}[ht] 
\begin{center}
\includegraphics[width=9cm]{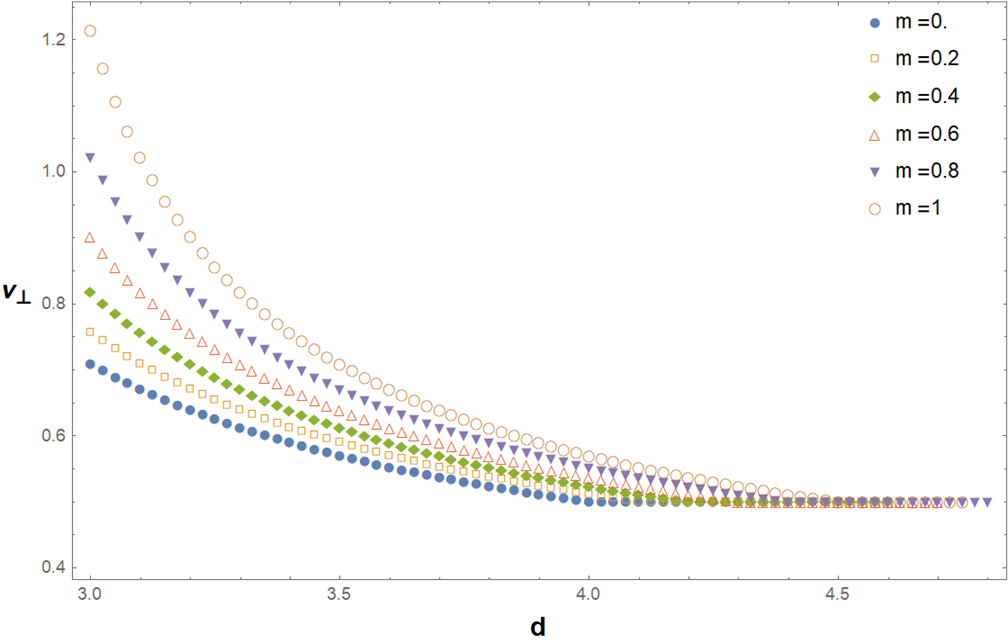}
\caption{The correlation length exponent $\nu_\perp$ for the $m$-axial Lifshitz point for $N=2$ plotted as a function of dimensionality $d$ for a sequence of values  of $m$. The plot demonstrates the shift of the upper critical dimension $d_u$ as well as the growing degree of divergence occurring upon increasing $m$. The curves are all related by translations in the horizontal direction (see the main text).  }
\label{d_dep}
\end{center} 
\end{figure} 
\begin{figure}[ht] 
\begin{center}
\includegraphics[width=9cm]{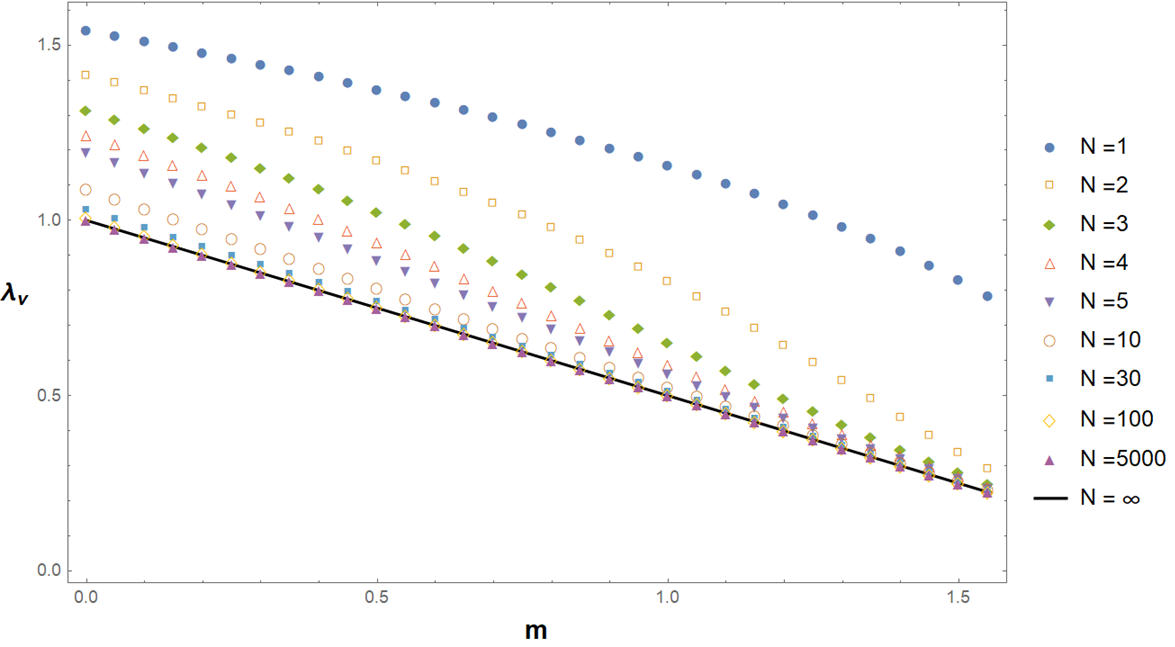}
\caption{The (inverse) correlation length exponent $\lambda_{\nu}=\nu_{\perp}^{-1}$ obtained within the LPA approximation plotted as a function of $m$ for $d=3$ and a sequence of values of $N$. The plot demonstrates in particular the convergence of the results towards the (exact) limit $\nu_{\perp}^{-1}\to d-2-\frac{m}{2}$ for $N\to\infty$. Interestingly, our results indicate significantly faster convergence for $m$ large.  }
\label{N_infty}
\end{center} 
\end{figure} 

Our approach correctly reproduces the $1/N\to 0$ as well as $\epsilon\to 0$ limits and is applicable to a broad range of parameters in the $(d,m,N)$ space. At the present truncation level it is however not sufficient to correctly address the limit of dimensionality $d$ approaching $d_L$, which is dominated by the neglected anomalous dimensions (see Sec.~IVC for an extension in this direction). Nonetheless our LPA data indicates a rapid growth of $\nu_\perp$ upon lowering $d$ towards $d_L$ (see e.g. Fig.~\ref{LPA_vs_eps}). 

\subsection{Constrained LPA' and the anomalous dimensions}
We now extend the truncation described in Sec.~IVB to account for the anomalous dimensions in the simplest conceivable way. This amounts to treating the quantities $Z_{\perp}$ and $Z_{||}$ in Eq.~(\ref{Gs}) as scale-dependent (but not field-dependent) quantities, while disregarding the term $\sim (\nabla_{||}\phi )^2$ (alike at the LPA level). The latter constitutes here an additional approximation. An analogous procedure for the case of isotropic $O(N)$ models is recognized as  ''LPA' '' and yields, for the $d=3$ XY or Heisenberg universality classes a somewhat overestimated value of the anomalous dimension $\eta$. For the present anisotropic situation (characterized by the effective dimensionality below 3) we may expect only a qualitative estimate of the values of  $\eta_\perp$ and $\eta_{||}$. Interestingly, we find nonetheless that the degree of violation of the correspondence discussed in Sec.~IVB upon including the anomalous dimensions is in fact very low in the physically interesting situations. 
Another interesting point concerns the sign of $\eta_{||}$. In this respect, for example in  $(d,m,N)=(3,1,3)$ the $1/N$ expansion (up to terms $\sim 1/N$) predicts\cite{Shpot_2005, Shpot_2012} a positive value in contrast to the $\epsilon$-expansion\cite{Shpot_2001} as well as the nonperturbative RG study of Ref.~(\onlinecite{Essafi_2012}). 

\begin{figure}[ht] 
\begin{center}
\includegraphics[width=9cm]{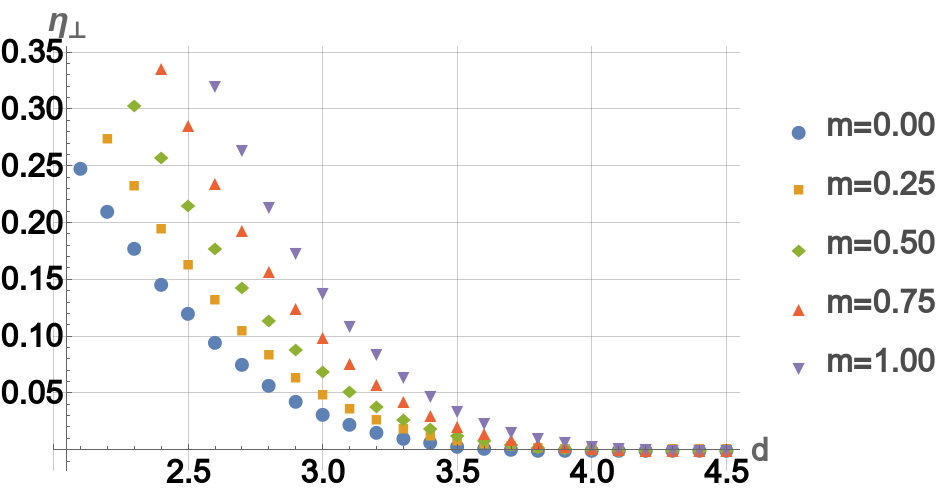}
\caption{The anomalous scaling dimension $\eta_\perp$ as function of $d$ for a sequence of values of $m$ and $N=1$. }
\label{eta_1}
\end{center} 
\end{figure}

\begin{figure}[ht] 
\begin{center}
\includegraphics[width=9cm]{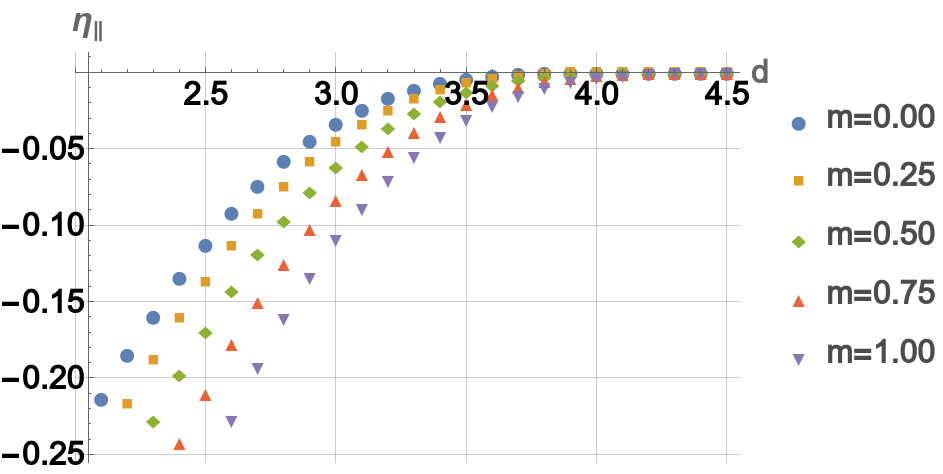}
\caption{ The anomalous scaling dimension $\eta_{||}$ as function of $d$ for a sequence of values of $m$ and $N=1$.}
\label{eta_2}
\end{center} 
\end{figure}

The running anomalous dimensions are related to the flowing $Z$-factors via\cite{Essafi_2012} $\eta_{\perp}=-\frac{1}{Z_\perp}\partial_t Z_{\perp}$ and $\eta_{||}=-\frac{1}{\theta Z_{||}}\partial_t Z_{||}$ with $\theta = \frac{2-\eta_{\perp}}{4-\eta_{||}}$ being the anisotropy exponent. The flow of the effective potential is derived along the line of Sec.~IVB. We obtain: 
\begin{align}
\partial_t \tu_k =& - \left(d-\frac{m}{2}-\frac{m}{4}(\eta_\perp-\eta_{||}) \right)\tilde{u}_k \nonumber \\ 
&-\left(2+\frac{m}{2}-d-\eta_\perp (1-\frac{m}{4})-\eta_{||}\frac{m}{4} \right)\trho \tu_k' \nonumber \\
&+\mathcal{V}_{d,m}\int_0^\infty dy y^{\frac{d}{2}-\frac{m}{4}-1}\times\nonumber \\ 
&\left[(2-\eta_\perp)r(y)-2yr'(y)\right] \left[\frac{1}{y+\tu_k'+2\trho\tu_k''+r(y)} +\frac{N-1}{y+\tu_k' +r(y)} \right] \nonumber \\
&+\mathcal{W}_{d,m}\int_0^\infty dy y^{\frac{d}{2}-\frac{m}{4}}\times\nonumber \\ 
&\left[\eta_\perp -\eta_{||}\right]r'(y) \left[\frac{1}{y+\tu_k'+2\trho\tu_k''+r(y)} +\frac{N-1}{y+\tu_k' +r(y)} \right]\;,
\label{LPA_prime_resc}
\end{align}
where 
\beq
\mathcal{W}_{d,m} =\frac{\mathcal{S}^{d-m-1}\mathcal{S}^{m-1}}{16(2\pi)^d}\mathcal{B}\left(\frac{d-m}{2},\frac{m}{4}+1\right)\;.
\eeq
The above flow equation for $u_k$ must be supplemented by the expressions for the running anomalous dimensions $\eta_\perp$ and $\eta_{||}$. These are evaluated along the line well described in literature (see e.g. Ref.~\onlinecite{Dupuis_2021}). By differentiating the Wetterich equation Eq.~(\ref{Wetterich}) twice, we obtain the flow of the two-point function $\Gamma^{(2)}$. Subsequently, by taking the second derivative with respect to momentum in the $\perp$ direction and the fourth derivative with respect to momentum in the $||$ direction evaluated at vanishing momentum, we extract the flow of $Z_\perp$ and $Z_{||}$, from which $\eta_\perp$ and $\eta_{||}$ follow. The resulting expressions (especially for $\eta_{||}$)  are very lengthy and we refrain from quoting them here.  
The physical anomalous scaling dimensions correspond to the fixed-point values of $\eta_\perp$ and $\eta_{||}$, which we extract numerically. The data presented below corresponds to results obtained with the PMS-optimized Wetterich cutoff. We note that for a range of $d$ and $m$ corresponding to low effective dimensionalities we were not able ot obtain a PMS value of $\alpha$, in which case we chose a value of $\alpha$ corresponding to a global extremum over a range of considered values. As representative results, in Figs.~\ref{eta_1} and \ref{eta_2}  we plot the obtained dependencies of $\eta_\perp$ and $\eta_{||}$ with fixed $N=1$, varying $d$ and $m$. Our numerical values are larger as compared to those resulting from the $\epsilon$ expansion which is probably due to both our truncation errors and the low order of the implemented $\epsilon$ expansion.  We point out that the sign of $\eta_{||}$ is negative in the entire region of parameters considered by us. As concerns the limit $m\to 0$, we observe  convergence of both $\eta_\perp$ and $\nu_\perp$ to the anticipated values corresponding to the standard $O(N)$ models. The quantity $\eta_{||}$ becomes a meaningless (redundant) parameter but does not vanish for $m\to 0$.      

\begin{figure}[ht] 
\begin{center}
\includegraphics[width=9cm]{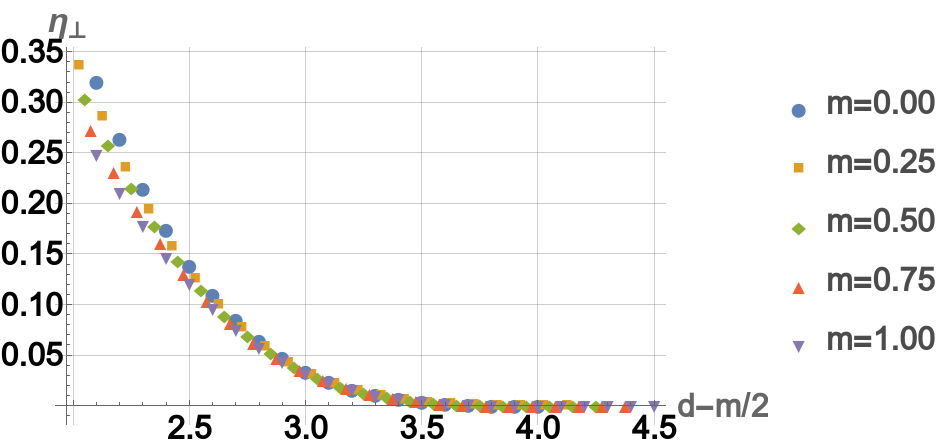}
\caption{The anomalous scaling dimension $\eta_\perp$ as function of $d_{eff}$ for a sequence of values of $m$ and $N=1$. The collapse of the curves indicates approximate fulfillment of the correspondence between the Lifshitz and $O(N)$-symmetric critical behavior also in presence of the anomalous dimensions.}
\label{eta_1_coll}
\end{center} 
\end{figure}

\begin{figure}[ht] 
\begin{center}
\includegraphics[width=9cm]{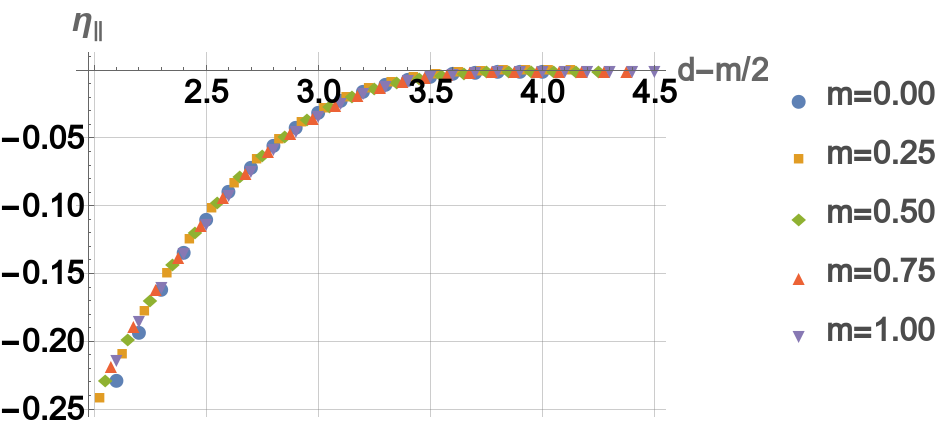}
\caption{ The anomalous scaling dimension $\eta_{||}$ as function of $d_{eff}$ for a sequence of values of $m$ and $N=1$. The collapse of the curves indicates approximate fulfillment of the correspondence between the Lifshitz and $O(N)$-symmetric critical behavior also in presence of the anomalous dimensions.}
\label{eta_2_coll}
\end{center} 
\end{figure}

We now investigate to which extent the relation between the Lifshitz point in $d$ dimensions and the $O(N)$ critical point in dimensionality $d_{eff}$ explored in Sec.~IVB becomes violated in presence of the anomalous dimensions. For this aim we plot the critical exponents as function of the effective dimensionality in Figs.~\ref{eta_1_coll} and ~\ref{eta_2_coll}. The collapse of the curves indicates a high level of  agreement with the picture demonstrated IVB at the LPA level for all of the critical exponents (including the anomalous dimensions).   
We by no means expect this equivalence to remain exact beyond the LPA approximation, however the level of numerical agreement is striking. 

We finally point out that the present LPA' level of approximation is the lowest possible allowing for capturing the anomalous dimensions. It would be very interesting to extend the present analysis by including $\rho$-dependencies as well as accounting for the  neglected $\sim (\nabla_{||}\phi )^2$ term. We relegate this to future studies.

\section{Summary}
In this paper we addressed the restrictions on the stability of the long-range ordered pair density wave (FFLO superfluid) states arising due to the presence of a thermal Lifshitz point as predicted by the mean-field theory. We pointed out that the occurrence of these phases in isotropic systems (such as ultracold atoms in continuum) is in fact completely excluded both in dimensionality $d=2$ and $d=3$, except for zero temperature. In consequence, the corresponding phase diagram should generically host a quantum Lifshitz point. This is no longer the case in systems exhibiting a unidirectional anisotropy, where a Lifshitz point at $T>0$ may be stable in $d=3$ (but not in $d=2$).   

The study of the FFLO superfluid Lifshitz point prompted us to readdress the Lifshitz critical behavior with arbitrary $d$, $m$ and $N$ from the point of view of functional renormalization group. We have found that at the approximation level of the local potential approximation (LPA), which amounts to disregarding the anomalous dimensions, the $m$-axial Lifshitz critical behavior is \emph{exactly} equivalent to that describing the standard $O(N)$-symmetric critical point in effective dimensionality $d_{eff}=d-m/2$. Our numerical analysis going beyond LPA level and accounting for $\eta_\perp$ and $\eta_{||}$ indicates that this relation is only mildly violated also in this case. In particular, we have found that the anomalous dimension $\eta_\perp$ with a high level of accuracy coincides with the value of the $\eta$ exponent of the corresponding $O(N)$ model in dimensionality reduced by $m/2$. We obtained negative values of $\eta_{||}$ for the entire range or scanned values of $(d, m, N)$. 

Our work opens avenues for future studies in at least two separate directions. On one hand, it would be interesting to explore thermodynamic and transport properties accompanying the vicinity of the fluctuation induced quantum Lifshitz point approaching it from finite $T$, in particular exploiting the interplay of order parameter and fermionic fluctuations. On the other hand, it might appear very fruitful to employ more sophisticated truncations of the Wetterich equation to further clarify the nature of the thermal Lifshitz points.      

\begin{acknowledgments}
We are grateful to Hans Werner Diehl, Dominique Mouhanna, Pierbiagio Pieri, and Mykola Shpot for useful correspondence and remarks on the content of the manuscript. P. J. thanks  Hiroyuki Yamase for numerous discussions on closely related topics. We acknowledge support from the Polish National Science Center via 2017/26/E/ST3/00211. 
\end{acknowledgments}


\begin{thebibliography}{73}%
\makeatletter
\providecommand \@ifxundefined [1]{%
 \@ifx{#1\undefined}
}%
\providecommand \@ifnum [1]{%
 \ifnum #1\expandafter \@firstoftwo
 \else \expandafter \@secondoftwo
 \fi
}%
\providecommand \@ifx [1]{%
 \ifx #1\expandafter \@firstoftwo
 \else \expandafter \@secondoftwo
 \fi
}%
\providecommand \natexlab [1]{#1}%
\providecommand \enquote  [1]{``#1''}%
\providecommand \bibnamefont  [1]{#1}%
\providecommand \bibfnamefont [1]{#1}%
\providecommand \citenamefont [1]{#1}%
\providecommand \href@noop [0]{\@secondoftwo}%
\providecommand \href [0]{\begingroup \@sanitize@url \@href}%
\providecommand \@href[1]{\@@startlink{#1}\@@href}%
\providecommand \@@href[1]{\endgroup#1\@@endlink}%
\providecommand \@sanitize@url [0]{\catcode `\\12\catcode `\$12\catcode
  `\&12\catcode `\#12\catcode `\^12\catcode `\_12\catcode `\%12\relax}%
\providecommand \@@startlink[1]{}%
\providecommand \@@endlink[0]{}%
\providecommand \url  [0]{\begingroup\@sanitize@url \@url }%
\providecommand \@url [1]{\endgroup\@href {#1}{\urlprefix }}%
\providecommand \urlprefix  [0]{URL }%
\providecommand \Eprint [0]{\href }%
\providecommand \doibase [0]{http://dx.doi.org/}%
\providecommand \selectlanguage [0]{\@gobble}%
\providecommand \bibinfo  [0]{\@secondoftwo}%
\providecommand \bibfield  [0]{\@secondoftwo}%
\providecommand \translation [1]{[#1]}%
\providecommand \BibitemOpen [0]{}%
\providecommand \bibitemStop [0]{}%
\providecommand \bibitemNoStop [0]{.\EOS\space}%
\providecommand \EOS [0]{\spacefactor3000\relax}%
\providecommand \BibitemShut  [1]{\csname bibitem#1\endcsname}%
\let\auto@bib@innerbib\@empty
\bibitem [{\citenamefont {Bloch}(2005)}]{Bloch_2005}%
  \BibitemOpen
  \bibfield  {author} {\bibinfo {author} {\bibfnamefont {I.}~\bibnamefont
  {Bloch}},\ }\href {\doibase 10.1038/nphys138} {\bibfield  {journal} {\bibinfo
   {journal} {Nature Physics}\ }\textbf {\bibinfo {volume} {1}},\ \bibinfo
  {pages} {23} (\bibinfo {year} {2005})}\BibitemShut {NoStop}%
\bibitem [{\citenamefont {Giorgini}\ \emph {et~al.}(2008)\citenamefont
  {Giorgini}, \citenamefont {Pitaevskii},\ and\ \citenamefont
  {Stringari}}]{giorgini_theory_2008}%
  \BibitemOpen
  \bibfield  {author} {\bibinfo {author} {\bibfnamefont {S.}~\bibnamefont
  {Giorgini}}, \bibinfo {author} {\bibfnamefont {L.~P.}\ \bibnamefont
  {Pitaevskii}}, \ and\ \bibinfo {author} {\bibfnamefont {S.}~\bibnamefont
  {Stringari}},\ }\href {\doibase 10.1103/RevModPhys.80.1215} {\bibfield
  {journal} {\bibinfo  {journal} {Rev. Mod. Phys.}\ }\textbf {\bibinfo {volume}
  {80}},\ \bibinfo {pages} {1215} (\bibinfo {year} {2008})}\BibitemShut
  {NoStop}%
\bibitem [{\citenamefont {Strinati}\ \emph {et~al.}(2018)\citenamefont
  {Strinati}, \citenamefont {Pieri}, \citenamefont {Rapke}, \citenamefont
  {Schuck},\ and\ \citenamefont {Urban}}]{Strinati_2018}%
  \BibitemOpen
  \bibfield  {author} {\bibinfo {author} {\bibfnamefont {G.~C.}\ \bibnamefont
  {Strinati}}, \bibinfo {author} {\bibfnamefont {P.}~\bibnamefont {Pieri}},
  \bibinfo {author} {\bibfnamefont {G.}~\bibnamefont {Rapke}}, \bibinfo
  {author} {\bibfnamefont {P.}~\bibnamefont {Schuck}}, \ and\ \bibinfo {author}
  {\bibfnamefont {M.}~\bibnamefont {Urban}},\ }\href {\doibase
  https://doi.org/10.1016/j.physrep.2018.02.004} {\bibfield  {journal}
  {\bibinfo  {journal} {Physics Reports}\ }\textbf {\bibinfo {volume} {738}},\
  \bibinfo {pages} {1 } (\bibinfo {year} {2018})}\BibitemShut {NoStop}%
\bibitem [{\citenamefont {Fulde}\ and\ \citenamefont
  {Ferrell}(1964)}]{fulde_superconductivity_1964}%
  \BibitemOpen
  \bibfield  {author} {\bibinfo {author} {\bibfnamefont {P.}~\bibnamefont
  {Fulde}}\ and\ \bibinfo {author} {\bibfnamefont {R.~A.}\ \bibnamefont
  {Ferrell}},\ }\href {\doibase 10.1103/PhysRev.135.A550} {\bibfield  {journal}
  {\bibinfo  {journal} {Phys. Rev.}\ }\textbf {\bibinfo {volume} {135}},\
  \bibinfo {pages} {A550} (\bibinfo {year} {1964})}\BibitemShut {NoStop}%
\bibitem [{\citenamefont {Larkin}\ and\ \citenamefont
  {Ovchinnikov}(1965)}]{larkin_nonuniform_1965}%
  \BibitemOpen
  \bibfield  {author} {\bibinfo {author} {\bibfnamefont {A.~I.}\ \bibnamefont
  {Larkin}}\ and\ \bibinfo {author} {\bibfnamefont {Y.~N.}\ \bibnamefont
  {Ovchinnikov}},\ }\href@noop {} {\bibfield  {journal} {\bibinfo  {journal}
  {Sov. Phys. JETP}\ }\textbf {\bibinfo {volume} {20}},\ \bibinfo {pages} {762}
  (\bibinfo {year} {1965})}\BibitemShut {NoStop}%
\bibitem [{\citenamefont {Agterberg}\ \emph {et~al.}(2020)\citenamefont
  {Agterberg}, \citenamefont {Davis}, \citenamefont {Edkins}, \citenamefont
  {Fradkin}, \citenamefont {Van~Harlingen}, \citenamefont {Kivelson},
  \citenamefont {Lee}, \citenamefont {Radzihovsky}, \citenamefont {Tranquada},\
  and\ \citenamefont {Wang}}]{Agterberg_2020}%
  \BibitemOpen
  \bibfield  {author} {\bibinfo {author} {\bibfnamefont {D.~F.}\ \bibnamefont
  {Agterberg}}, \bibinfo {author} {\bibfnamefont {J.~S.}\ \bibnamefont
  {Davis}}, \bibinfo {author} {\bibfnamefont {S.~D.}\ \bibnamefont {Edkins}},
  \bibinfo {author} {\bibfnamefont {E.}~\bibnamefont {Fradkin}}, \bibinfo
  {author} {\bibfnamefont {D.~J.}\ \bibnamefont {Van~Harlingen}}, \bibinfo
  {author} {\bibfnamefont {S.~A.}\ \bibnamefont {Kivelson}}, \bibinfo {author}
  {\bibfnamefont {P.~A.}\ \bibnamefont {Lee}}, \bibinfo {author} {\bibfnamefont
  {L.}~\bibnamefont {Radzihovsky}}, \bibinfo {author} {\bibfnamefont {J.~M.}\
  \bibnamefont {Tranquada}}, \ and\ \bibinfo {author} {\bibfnamefont
  {Y.}~\bibnamefont {Wang}},\ }\href@noop {} {\bibfield  {journal} {\bibinfo
  {journal} {Annual Review of Condensed Matter Physics}\ }\textbf {\bibinfo
  {volume} {11}},\ \bibinfo {pages} {231} (\bibinfo {year} {2020})}\BibitemShut
  {NoStop}%
\bibitem [{\citenamefont {He}\ \emph {et~al.}(2006)\citenamefont {He},
  \citenamefont {Jin},\ and\ \citenamefont {Zhuang}}]{He_2006}%
  \BibitemOpen
  \bibfield  {author} {\bibinfo {author} {\bibfnamefont {L.}~\bibnamefont
  {He}}, \bibinfo {author} {\bibfnamefont {M.}~\bibnamefont {Jin}}, \ and\
  \bibinfo {author} {\bibfnamefont {P.}~\bibnamefont {Zhuang}},\ }\href
  {\doibase 10.1103/PhysRevB.73.214527} {\bibfield  {journal} {\bibinfo
  {journal} {Phys. Rev. B}\ }\textbf {\bibinfo {volume} {73}},\ \bibinfo
  {pages} {214527} (\bibinfo {year} {2006})}\BibitemShut {NoStop}%
\bibitem [{\citenamefont {Gubbels}\ \emph {et~al.}(2009)\citenamefont
  {Gubbels}, \citenamefont {Baarsma},\ and\ \citenamefont
  {Stoof}}]{Gubbels_2009}%
  \BibitemOpen
  \bibfield  {author} {\bibinfo {author} {\bibfnamefont {K.~B.}\ \bibnamefont
  {Gubbels}}, \bibinfo {author} {\bibfnamefont {J.~E.}\ \bibnamefont
  {Baarsma}}, \ and\ \bibinfo {author} {\bibfnamefont {H.~T.~C.}\ \bibnamefont
  {Stoof}},\ }\href {\doibase 10.1103/PhysRevLett.103.195301} {\bibfield
  {journal} {\bibinfo  {journal} {Phys. Rev. Lett.}\ }\textbf {\bibinfo
  {volume} {103}},\ \bibinfo {pages} {195301} (\bibinfo {year}
  {2009})}\BibitemShut {NoStop}%
\bibitem [{\citenamefont {Baarsma}\ \emph {et~al.}(2010)\citenamefont
  {Baarsma}, \citenamefont {Gubbels},\ and\ \citenamefont
  {Stoof}}]{Baarsma_2010}%
  \BibitemOpen
  \bibfield  {author} {\bibinfo {author} {\bibfnamefont {J.~E.}\ \bibnamefont
  {Baarsma}}, \bibinfo {author} {\bibfnamefont {K.~B.}\ \bibnamefont
  {Gubbels}}, \ and\ \bibinfo {author} {\bibfnamefont {H.~T.~C.}\ \bibnamefont
  {Stoof}},\ }\href {\doibase 10.1103/PhysRevA.82.013624} {\bibfield  {journal}
  {\bibinfo  {journal} {Phys. Rev. A}\ }\textbf {\bibinfo {volume} {82}},\
  \bibinfo {pages} {013624} (\bibinfo {year} {2010})}\BibitemShut {NoStop}%
\bibitem [{\citenamefont {Radzihovsky}\ and\ \citenamefont
  {Sheehy}(2010)}]{radzihovsky_imbalanced_2010}%
  \BibitemOpen
  \bibfield  {author} {\bibinfo {author} {\bibfnamefont {L.}~\bibnamefont
  {Radzihovsky}}\ and\ \bibinfo {author} {\bibfnamefont {D.~E.}\ \bibnamefont
  {Sheehy}},\ }\href {\doibase 10.1088/0034-4885/73/7/076501} {\bibfield
  {journal} {\bibinfo  {journal} {Rep. Prog. Phys.}\ }\textbf {\bibinfo
  {volume} {73}},\ \bibinfo {pages} {076501} (\bibinfo {year}
  {2010})}\BibitemShut {NoStop}%
\bibitem [{\citenamefont {Cai}\ \emph {et~al.}(2011)\citenamefont {Cai},
  \citenamefont {Wang},\ and\ \citenamefont {Wu}}]{Cai_2011}%
  \BibitemOpen
  \bibfield  {author} {\bibinfo {author} {\bibfnamefont {Z.}~\bibnamefont
  {Cai}}, \bibinfo {author} {\bibfnamefont {Y.}~\bibnamefont {Wang}}, \ and\
  \bibinfo {author} {\bibfnamefont {C.}~\bibnamefont {Wu}},\ }\href {\doibase
  10.1103/PhysRevA.83.063621} {\bibfield  {journal} {\bibinfo  {journal} {Phys.
  Rev. A}\ }\textbf {\bibinfo {volume} {83}},\ \bibinfo {pages} {063621}
  (\bibinfo {year} {2011})}\BibitemShut {NoStop}%
\bibitem [{\citenamefont {Baarsma}\ and\ \citenamefont
  {Stoof}(2013)}]{Baarsma_2013}%
  \BibitemOpen
  \bibfield  {author} {\bibinfo {author} {\bibfnamefont {J.~E.}\ \bibnamefont
  {Baarsma}}\ and\ \bibinfo {author} {\bibfnamefont {H.~T.~C.}\ \bibnamefont
  {Stoof}},\ }\href {\doibase 10.1103/PhysRevA.87.063612} {\bibfield  {journal}
  {\bibinfo  {journal} {Phys. Rev. A}\ }\textbf {\bibinfo {volume} {87}},\
  \bibinfo {pages} {063612} (\bibinfo {year} {2013})}\BibitemShut {NoStop}%
\bibitem [{\citenamefont {Roscher}\ \emph {et~al.}(2015)\citenamefont
  {Roscher}, \citenamefont {Braun},\ and\ \citenamefont {Drut}}]{Rosher_2015}%
  \BibitemOpen
  \bibfield  {author} {\bibinfo {author} {\bibfnamefont {D.}~\bibnamefont
  {Roscher}}, \bibinfo {author} {\bibfnamefont {J.}~\bibnamefont {Braun}}, \
  and\ \bibinfo {author} {\bibfnamefont {J.~E.}\ \bibnamefont {Drut}},\ }\href
  {\doibase 10.1103/PhysRevA.91.053611} {\bibfield  {journal} {\bibinfo
  {journal} {Phys. Rev. A}\ }\textbf {\bibinfo {volume} {91}},\ \bibinfo
  {pages} {053611} (\bibinfo {year} {2015})}\BibitemShut {NoStop}%
\bibitem [{\citenamefont {Karmakar}\ and\ \citenamefont
  {Majumdar}(2016)}]{Karmakar_2016}%
  \BibitemOpen
  \bibfield  {author} {\bibinfo {author} {\bibfnamefont {M.}~\bibnamefont
  {Karmakar}}\ and\ \bibinfo {author} {\bibfnamefont {P.}~\bibnamefont
  {Majumdar}},\ }\href {\doibase 10.1103/PhysRevA.93.053609} {\bibfield
  {journal} {\bibinfo  {journal} {Phys. Rev. A}\ }\textbf {\bibinfo {volume}
  {93}},\ \bibinfo {pages} {053609} (\bibinfo {year} {2016})}\BibitemShut
  {NoStop}%
\bibitem [{\citenamefont {Kinnunen}\ \emph {et~al.}(2018)\citenamefont
  {Kinnunen}, \citenamefont {Baarsma}, \citenamefont {Martikainen},\ and\
  \citenamefont {Torma}}]{Kinnunen_2018}%
  \BibitemOpen
  \bibfield  {author} {\bibinfo {author} {\bibfnamefont {J.~J.}\ \bibnamefont
  {Kinnunen}}, \bibinfo {author} {\bibfnamefont {J.}~\bibnamefont {Baarsma}},
  \bibinfo {author} {\bibfnamefont {J.-P.}\ \bibnamefont {Martikainen}}, \ and\
  \bibinfo {author} {\bibfnamefont {P.}~\bibnamefont {Torma}},\ }\href
  {http://iopscience.iop.org/10.1088/1361-6633/aaa4ad} {\bibfield  {journal}
  {\bibinfo  {journal} {Reports on Progress in Physics}\ } (\bibinfo {year}
  {2018})}\BibitemShut {NoStop}%
\bibitem [{\citenamefont {Pini}\ \emph
  {et~al.}(2021{\natexlab{a}})\citenamefont {Pini}, \citenamefont {Pieri},
  \citenamefont {Grimm},\ and\ \citenamefont {Strinati}}]{Pini_2021_2}%
  \BibitemOpen
  \bibfield  {author} {\bibinfo {author} {\bibfnamefont {M.}~\bibnamefont
  {Pini}}, \bibinfo {author} {\bibfnamefont {P.}~\bibnamefont {Pieri}},
  \bibinfo {author} {\bibfnamefont {R.}~\bibnamefont {Grimm}}, \ and\ \bibinfo
  {author} {\bibfnamefont {G.~C.}\ \bibnamefont {Strinati}},\ }\href {\doibase
  10.1103/PhysRevA.103.023314} {\bibfield  {journal} {\bibinfo  {journal}
  {Phys. Rev. A}\ }\textbf {\bibinfo {volume} {103}},\ \bibinfo {pages}
  {023314} (\bibinfo {year} {2021}{\natexlab{a}})}\BibitemShut {NoStop}%
\bibitem [{\citenamefont {Rammelm\"uller}\ \emph {et~al.}(2021)\citenamefont
  {Rammelm\"uller}, \citenamefont {Hou}, \citenamefont {Drut},\ and\
  \citenamefont {Braun}}]{Rammelmuller_2021}%
  \BibitemOpen
  \bibfield  {author} {\bibinfo {author} {\bibfnamefont {L.}~\bibnamefont
  {Rammelm\"uller}}, \bibinfo {author} {\bibfnamefont {Y.}~\bibnamefont {Hou}},
  \bibinfo {author} {\bibfnamefont {J.~E.}\ \bibnamefont {Drut}}, \ and\
  \bibinfo {author} {\bibfnamefont {J.}~\bibnamefont {Braun}},\ }\href
  {\doibase 10.1103/PhysRevA.103.043330} {\bibfield  {journal} {\bibinfo
  {journal} {Phys. Rev. A}\ }\textbf {\bibinfo {volume} {103}},\ \bibinfo
  {pages} {043330} (\bibinfo {year} {2021})}\BibitemShut {NoStop}%
\bibitem [{\citenamefont {Shimahara}(1998)}]{Shimahara_1998}%
  \BibitemOpen
  \bibfield  {author} {\bibinfo {author} {\bibfnamefont {H.}~\bibnamefont
  {Shimahara}},\ }\href@noop {} {\bibfield  {journal} {\bibinfo  {journal}
  {Journal of the Physical Society of Japan}\ }\textbf {\bibinfo {volume}
  {67}},\ \bibinfo {pages} {1872} (\bibinfo {year} {1998})}\BibitemShut
  {NoStop}%
\bibitem [{\citenamefont {Samokhin}(2010)}]{Samokhin_2010}%
  \BibitemOpen
  \bibfield  {author} {\bibinfo {author} {\bibfnamefont {K.~V.}\ \bibnamefont
  {Samokhin}},\ }\href {\doibase 10.1103/PhysRevB.81.224507} {\bibfield
  {journal} {\bibinfo  {journal} {Phys. Rev. B}\ }\textbf {\bibinfo {volume}
  {81}},\ \bibinfo {pages} {224507} (\bibinfo {year} {2010})}\BibitemShut
  {NoStop}%
\bibitem [{\citenamefont {Radzihovsky}\ and\ \citenamefont
  {Vishwanath}(2009)}]{Radzihovsky_2009}%
  \BibitemOpen
  \bibfield  {author} {\bibinfo {author} {\bibfnamefont {L.}~\bibnamefont
  {Radzihovsky}}\ and\ \bibinfo {author} {\bibfnamefont {A.}~\bibnamefont
  {Vishwanath}},\ }\href {\doibase 10.1103/PhysRevLett.103.010404} {\bibfield
  {journal} {\bibinfo  {journal} {Phys. Rev. Lett.}\ }\textbf {\bibinfo
  {volume} {103}},\ \bibinfo {pages} {010404} (\bibinfo {year}
  {2009})}\BibitemShut {NoStop}%
\bibitem [{\citenamefont {Radzihovsky}(2011)}]{Radzihovsky_2011}%
  \BibitemOpen
  \bibfield  {author} {\bibinfo {author} {\bibfnamefont {L.}~\bibnamefont
  {Radzihovsky}},\ }\href {\doibase 10.1103/PhysRevA.84.023611} {\bibfield
  {journal} {\bibinfo  {journal} {Phys. Rev. A}\ }\textbf {\bibinfo {volume}
  {84}},\ \bibinfo {pages} {023611} (\bibinfo {year} {2011})}\BibitemShut
  {NoStop}%
\bibitem [{\citenamefont {Yin}\ \emph {et~al.}(2014)\citenamefont {Yin},
  \citenamefont {Martikainen},\ and\ \citenamefont {T\"orm\"a}}]{Yin_2014}%
  \BibitemOpen
  \bibfield  {author} {\bibinfo {author} {\bibfnamefont {S.}~\bibnamefont
  {Yin}}, \bibinfo {author} {\bibfnamefont {J.-P.}\ \bibnamefont
  {Martikainen}}, \ and\ \bibinfo {author} {\bibfnamefont {P.}~\bibnamefont
  {T\"orm\"a}},\ }\href {\doibase 10.1103/PhysRevB.89.014507} {\bibfield
  {journal} {\bibinfo  {journal} {Phys. Rev. B}\ }\textbf {\bibinfo {volume}
  {89}},\ \bibinfo {pages} {014507} (\bibinfo {year} {2014})}\BibitemShut
  {NoStop}%
\bibitem [{\citenamefont {Jakubczyk}(2017)}]{Jakubczyk_2017}%
  \BibitemOpen
  \bibfield  {author} {\bibinfo {author} {\bibfnamefont {P.}~\bibnamefont
  {Jakubczyk}},\ }\href {\doibase 10.1103/PhysRevA.95.063626} {\bibfield
  {journal} {\bibinfo  {journal} {Phys. Rev. A}\ }\textbf {\bibinfo {volume}
  {95}},\ \bibinfo {pages} {063626} (\bibinfo {year} {2017})}\BibitemShut
  {NoStop}%
\bibitem [{\citenamefont {Wang}\ \emph {et~al.}(2018)\citenamefont {Wang},
  \citenamefont {Che}, \citenamefont {Zhang},\ and\ \citenamefont
  {Chen}}]{Wang_2018}%
  \BibitemOpen
  \bibfield  {author} {\bibinfo {author} {\bibfnamefont {J.}~\bibnamefont
  {Wang}}, \bibinfo {author} {\bibfnamefont {Y.}~\bibnamefont {Che}}, \bibinfo
  {author} {\bibfnamefont {L.}~\bibnamefont {Zhang}}, \ and\ \bibinfo {author}
  {\bibfnamefont {Q.}~\bibnamefont {Chen}},\ }\href {\doibase
  10.1103/PhysRevB.97.134513} {\bibfield  {journal} {\bibinfo  {journal} {Phys.
  Rev. B}\ }\textbf {\bibinfo {volume} {97}},\ \bibinfo {pages} {134513}
  (\bibinfo {year} {2018})}\BibitemShut {NoStop}%
\bibitem [{\citenamefont {Lutchyn}\ \emph {et~al.}(2011)\citenamefont
  {Lutchyn}, \citenamefont {Dzero},\ and\ \citenamefont
  {Yakovenko}}]{Lutchyn_2011}%
  \BibitemOpen
  \bibfield  {author} {\bibinfo {author} {\bibfnamefont {R.~M.}\ \bibnamefont
  {Lutchyn}}, \bibinfo {author} {\bibfnamefont {M.}~\bibnamefont {Dzero}}, \
  and\ \bibinfo {author} {\bibfnamefont {V.~M.}\ \bibnamefont {Yakovenko}},\
  }\href {\doibase 10.1103/PhysRevA.84.033609} {\bibfield  {journal} {\bibinfo
  {journal} {Phys. Rev. A}\ }\textbf {\bibinfo {volume} {84}},\ \bibinfo
  {pages} {033609} (\bibinfo {year} {2011})}\BibitemShut {NoStop}%
\bibitem [{\citenamefont {Revelle}\ \emph {et~al.}(2016)\citenamefont
  {Revelle}, \citenamefont {Fry}, \citenamefont {Olsen},\ and\ \citenamefont
  {Hulet}}]{Revelle_2016}%
  \BibitemOpen
  \bibfield  {author} {\bibinfo {author} {\bibfnamefont {M.~C.}\ \bibnamefont
  {Revelle}}, \bibinfo {author} {\bibfnamefont {J.~A.}\ \bibnamefont {Fry}},
  \bibinfo {author} {\bibfnamefont {B.~A.}\ \bibnamefont {Olsen}}, \ and\
  \bibinfo {author} {\bibfnamefont {R.~G.}\ \bibnamefont {Hulet}},\ }\href
  {\doibase 10.1103/PhysRevLett.117.235301} {\bibfield  {journal} {\bibinfo
  {journal} {Phys. Rev. Lett.}\ }\textbf {\bibinfo {volume} {117}},\ \bibinfo
  {pages} {235301} (\bibinfo {year} {2016})}\BibitemShut {NoStop}%
\bibitem [{\citenamefont {Sundar}\ \emph {et~al.}(2020)\citenamefont {Sundar},
  \citenamefont {Fry}, \citenamefont {Revelle}, \citenamefont {Hulet},\ and\
  \citenamefont {Hazzard}}]{Sundar_2020}%
  \BibitemOpen
  \bibfield  {author} {\bibinfo {author} {\bibfnamefont {B.}~\bibnamefont
  {Sundar}}, \bibinfo {author} {\bibfnamefont {J.~A.}\ \bibnamefont {Fry}},
  \bibinfo {author} {\bibfnamefont {M.~C.}\ \bibnamefont {Revelle}}, \bibinfo
  {author} {\bibfnamefont {R.~G.}\ \bibnamefont {Hulet}}, \ and\ \bibinfo
  {author} {\bibfnamefont {K.~R.~A.}\ \bibnamefont {Hazzard}},\ }\href
  {\doibase 10.1103/PhysRevA.102.033311} {\bibfield  {journal} {\bibinfo
  {journal} {Phys. Rev. A}\ }\textbf {\bibinfo {volume} {102}},\ \bibinfo
  {pages} {033311} (\bibinfo {year} {2020})}\BibitemShut {NoStop}%
\bibitem [{\citenamefont {Zdybel}\ and\ \citenamefont
  {Jakubczyk}(2020)}]{Zdybel_2020}%
  \BibitemOpen
  \bibfield  {author} {\bibinfo {author} {\bibfnamefont {P.}~\bibnamefont
  {Zdybel}}\ and\ \bibinfo {author} {\bibfnamefont {P.}~\bibnamefont
  {Jakubczyk}},\ }\href {\doibase 10.1103/PhysRevResearch.2.033486} {\bibfield
  {journal} {\bibinfo  {journal} {Phys. Rev. Research}\ }\textbf {\bibinfo
  {volume} {2}},\ \bibinfo {pages} {033486} (\bibinfo {year}
  {2020})}\BibitemShut {NoStop}%
\bibitem [{\citenamefont {Uji}\ \emph {et~al.}(2012)\citenamefont {Uji},
  \citenamefont {Kodama}, \citenamefont {Sugii}, \citenamefont {Terashima},
  \citenamefont {Takahide}, \citenamefont {Kurita}, \citenamefont {Tsuchiya},
  \citenamefont {Kimata}, \citenamefont {Kobayashi}, \citenamefont {Zhou},\
  and\ \citenamefont {Kobayashi}}]{Uji_2012}%
  \BibitemOpen
  \bibfield  {author} {\bibinfo {author} {\bibfnamefont {S.}~\bibnamefont
  {Uji}}, \bibinfo {author} {\bibfnamefont {K.}~\bibnamefont {Kodama}},
  \bibinfo {author} {\bibfnamefont {K.}~\bibnamefont {Sugii}}, \bibinfo
  {author} {\bibfnamefont {T.}~\bibnamefont {Terashima}}, \bibinfo {author}
  {\bibfnamefont {Y.}~\bibnamefont {Takahide}}, \bibinfo {author}
  {\bibfnamefont {N.}~\bibnamefont {Kurita}}, \bibinfo {author} {\bibfnamefont
  {S.}~\bibnamefont {Tsuchiya}}, \bibinfo {author} {\bibfnamefont
  {M.}~\bibnamefont {Kimata}}, \bibinfo {author} {\bibfnamefont
  {A.}~\bibnamefont {Kobayashi}}, \bibinfo {author} {\bibfnamefont
  {B.}~\bibnamefont {Zhou}}, \ and\ \bibinfo {author} {\bibfnamefont
  {H.}~\bibnamefont {Kobayashi}},\ }\href {\doibase 10.1103/PhysRevB.85.174530}
  {\bibfield  {journal} {\bibinfo  {journal} {Phys. Rev. B}\ }\textbf {\bibinfo
  {volume} {85}},\ \bibinfo {pages} {174530} (\bibinfo {year}
  {2012})}\BibitemShut {NoStop}%
\bibitem [{\citenamefont {Uji}\ \emph {et~al.}(2013)\citenamefont {Uji},
  \citenamefont {Kodama}, \citenamefont {Sugii}, \citenamefont {Terashima},
  \citenamefont {Yamaguchi}, \citenamefont {Kurita}, \citenamefont {Tsuchiya},
  \citenamefont {Kimata}, \citenamefont {Konoike}, \citenamefont {Kobayashi},
  \citenamefont {Zhou},\ and\ \citenamefont {Kobayashi}}]{Uji_2013}%
  \BibitemOpen
  \bibfield  {author} {\bibinfo {author} {\bibfnamefont {S.}~\bibnamefont
  {Uji}}, \bibinfo {author} {\bibfnamefont {K.}~\bibnamefont {Kodama}},
  \bibinfo {author} {\bibfnamefont {K.}~\bibnamefont {Sugii}}, \bibinfo
  {author} {\bibfnamefont {T.}~\bibnamefont {Terashima}}, \bibinfo {author}
  {\bibfnamefont {T.}~\bibnamefont {Yamaguchi}}, \bibinfo {author}
  {\bibfnamefont {N.}~\bibnamefont {Kurita}}, \bibinfo {author} {\bibfnamefont
  {S.}~\bibnamefont {Tsuchiya}}, \bibinfo {author} {\bibfnamefont
  {M.}~\bibnamefont {Kimata}}, \bibinfo {author} {\bibfnamefont
  {T.}~\bibnamefont {Konoike}}, \bibinfo {author} {\bibfnamefont
  {A.}~\bibnamefont {Kobayashi}}, \bibinfo {author} {\bibfnamefont
  {B.}~\bibnamefont {Zhou}}, \ and\ \bibinfo {author} {\bibfnamefont
  {H.}~\bibnamefont {Kobayashi}},\ }\href {\doibase 10.7566/JPSJ.82.034715}
  {\bibfield  {journal} {\bibinfo  {journal} {Journal of the Physical Society
  of Japan}\ }\textbf {\bibinfo {volume} {82}},\ \bibinfo {pages} {034715}
  (\bibinfo {year} {2013})}\BibitemShut {NoStop}%
\bibitem [{\citenamefont {Tsuchiya}\ \emph {et~al.}(2015)\citenamefont
  {Tsuchiya}, \citenamefont {Yamada}, \citenamefont {Sugii}, \citenamefont
  {Graf}, \citenamefont {Brooks}, \citenamefont {Terashima},\ and\
  \citenamefont {Uji}}]{Tsuchiya_2015}%
  \BibitemOpen
  \bibfield  {author} {\bibinfo {author} {\bibfnamefont {S.}~\bibnamefont
  {Tsuchiya}}, \bibinfo {author} {\bibfnamefont {J.-i.}\ \bibnamefont
  {Yamada}}, \bibinfo {author} {\bibfnamefont {K.}~\bibnamefont {Sugii}},
  \bibinfo {author} {\bibfnamefont {D.}~\bibnamefont {Graf}}, \bibinfo {author}
  {\bibfnamefont {J.~S.}\ \bibnamefont {Brooks}}, \bibinfo {author}
  {\bibfnamefont {T.}~\bibnamefont {Terashima}}, \ and\ \bibinfo {author}
  {\bibfnamefont {S.}~\bibnamefont {Uji}},\ }\href {\doibase
  10.7566/JPSJ.84.034703} {\bibfield  {journal} {\bibinfo  {journal} {Journal
  of the Physical Society of Japan}\ }\textbf {\bibinfo {volume} {84}},\
  \bibinfo {pages} {034703} (\bibinfo {year} {2015})}\BibitemShut {NoStop}%
\bibitem [{\citenamefont {Koutroulakis}\ \emph {et~al.}(2016)\citenamefont
  {Koutroulakis}, \citenamefont {K\"uhne}, \citenamefont {Schlueter},
  \citenamefont {Wosnitza},\ and\ \citenamefont {Brown}}]{Koutroulakis_2016}%
  \BibitemOpen
  \bibfield  {author} {\bibinfo {author} {\bibfnamefont {G.}~\bibnamefont
  {Koutroulakis}}, \bibinfo {author} {\bibfnamefont {H.}~\bibnamefont
  {K\"uhne}}, \bibinfo {author} {\bibfnamefont {J.~A.}\ \bibnamefont
  {Schlueter}}, \bibinfo {author} {\bibfnamefont {J.}~\bibnamefont {Wosnitza}},
  \ and\ \bibinfo {author} {\bibfnamefont {S.~E.}\ \bibnamefont {Brown}},\
  }\href {\doibase 10.1103/PhysRevLett.116.067003} {\bibfield  {journal}
  {\bibinfo  {journal} {Phys. Rev. Lett.}\ }\textbf {\bibinfo {volume} {116}},\
  \bibinfo {pages} {067003} (\bibinfo {year} {2016})}\BibitemShut {NoStop}%
\bibitem [{\citenamefont {Cho}\ \emph {et~al.}(2021)\citenamefont {Cho},
  \citenamefont {Lyu}, \citenamefont {Ng}, \citenamefont {He}, \citenamefont
  {Lo}, \citenamefont {Chareev}, \citenamefont {Abdel-Baset}, \citenamefont
  {Abdel-Hafiez},\ and\ \citenamefont {Lortz}}]{Cho_2021}%
  \BibitemOpen
  \bibfield  {author} {\bibinfo {author} {\bibfnamefont {C.-w.}\ \bibnamefont
  {Cho}}, \bibinfo {author} {\bibfnamefont {J.}~\bibnamefont {Lyu}}, \bibinfo
  {author} {\bibfnamefont {C.~Y.}\ \bibnamefont {Ng}}, \bibinfo {author}
  {\bibfnamefont {J.~J.}\ \bibnamefont {He}}, \bibinfo {author} {\bibfnamefont
  {K.~T.}\ \bibnamefont {Lo}}, \bibinfo {author} {\bibfnamefont
  {D.}~\bibnamefont {Chareev}}, \bibinfo {author} {\bibfnamefont {T.~A.}\
  \bibnamefont {Abdel-Baset}}, \bibinfo {author} {\bibfnamefont
  {M.}~\bibnamefont {Abdel-Hafiez}}, \ and\ \bibinfo {author} {\bibfnamefont
  {R.}~\bibnamefont {Lortz}},\ }\href {\doibase 10.1038/s41467-021-23976-2}
  {\bibfield  {journal} {\bibinfo  {journal} {Nature Communications}\ }\textbf
  {\bibinfo {volume} {12}},\ \bibinfo {pages} {3676} (\bibinfo {year}
  {2021})}\BibitemShut {NoStop}%
\bibitem [{\citenamefont {Mayaffre}\ \emph {et~al.}(2014)\citenamefont
  {Mayaffre}, \citenamefont {Kr{\"a}mer}, \citenamefont {Horvati{\'{c}}},
  \citenamefont {Berthier}, \citenamefont {Miyagawa}, \citenamefont {Kanoda},\
  and\ \citenamefont {Mitrovi{\'{c}}}}]{Mayaffre_2014}%
  \BibitemOpen
  \bibfield  {author} {\bibinfo {author} {\bibfnamefont {H.}~\bibnamefont
  {Mayaffre}}, \bibinfo {author} {\bibfnamefont {S.}~\bibnamefont
  {Kr{\"a}mer}}, \bibinfo {author} {\bibfnamefont {M.}~\bibnamefont
  {Horvati{\'{c}}}}, \bibinfo {author} {\bibfnamefont {C.}~\bibnamefont
  {Berthier}}, \bibinfo {author} {\bibfnamefont {K.}~\bibnamefont {Miyagawa}},
  \bibinfo {author} {\bibfnamefont {K.}~\bibnamefont {Kanoda}}, \ and\ \bibinfo
  {author} {\bibfnamefont {V.~F.}\ \bibnamefont {Mitrovi{\'{c}}}},\ }\href
  {\doibase 10.1038/nphys3121} {\bibfield  {journal} {\bibinfo  {journal}
  {Nature Physics}\ }\textbf {\bibinfo {volume} {10}},\ \bibinfo {pages} {928}
  (\bibinfo {year} {2014})}\BibitemShut {NoStop}%
\bibitem [{\citenamefont {Piazza}\ \emph {et~al.}(2016)\citenamefont {Piazza},
  \citenamefont {Zwerger},\ and\ \citenamefont {Strack}}]{Piazza_2016}%
  \BibitemOpen
  \bibfield  {author} {\bibinfo {author} {\bibfnamefont {F.}~\bibnamefont
  {Piazza}}, \bibinfo {author} {\bibfnamefont {W.}~\bibnamefont {Zwerger}}, \
  and\ \bibinfo {author} {\bibfnamefont {P.}~\bibnamefont {Strack}},\ }\href
  {\doibase 10.1103/PhysRevB.93.085112} {\bibfield  {journal} {\bibinfo
  {journal} {Phys. Rev. B}\ }\textbf {\bibinfo {volume} {93}},\ \bibinfo
  {pages} {085112} (\bibinfo {year} {2016})}\BibitemShut {NoStop}%
\bibitem [{\citenamefont {Strack}\ and\ \citenamefont
  {Jakubczyk}(2014)}]{Strack_2014}%
  \BibitemOpen
  \bibfield  {author} {\bibinfo {author} {\bibfnamefont {P.}~\bibnamefont
  {Strack}}\ and\ \bibinfo {author} {\bibfnamefont {P.}~\bibnamefont
  {Jakubczyk}},\ }\href {\doibase 10.1103/PhysRevX.4.021012} {\bibfield
  {journal} {\bibinfo  {journal} {Phys. Rev. X}\ }\textbf {\bibinfo {volume}
  {4}},\ \bibinfo {pages} {021012} (\bibinfo {year} {2014})}\BibitemShut
  {NoStop}%
\bibitem [{\citenamefont {Zdybel}\ and\ \citenamefont
  {Jakubczyk}(2018)}]{Zdybel_2018}%
  \BibitemOpen
  \bibfield  {author} {\bibinfo {author} {\bibfnamefont {P.}~\bibnamefont
  {Zdybel}}\ and\ \bibinfo {author} {\bibfnamefont {P.}~\bibnamefont
  {Jakubczyk}},\ }\href {\doibase 10.1088/1361-648x/aacc00} {\bibfield
  {journal} {\bibinfo  {journal} {Journal of Physics: Condensed Matter}\
  }\textbf {\bibinfo {volume} {30}},\ \bibinfo {pages} {305604} (\bibinfo
  {year} {2018})}\BibitemShut {NoStop}%
\bibitem [{\citenamefont {Zdybel}\ and\ \citenamefont
  {Jakubczyk}(2019)}]{Zdybel_2019}%
  \BibitemOpen
  \bibfield  {author} {\bibinfo {author} {\bibfnamefont {P.}~\bibnamefont
  {Zdybel}}\ and\ \bibinfo {author} {\bibfnamefont {P.}~\bibnamefont
  {Jakubczyk}},\ }\href {\doibase 10.1103/PhysRevA.100.053622} {\bibfield
  {journal} {\bibinfo  {journal} {Phys. Rev. A}\ }\textbf {\bibinfo {volume}
  {100}},\ \bibinfo {pages} {053622} (\bibinfo {year} {2019})}\BibitemShut
  {NoStop}%
\bibitem [{\citenamefont {Nagaosa}(1999)}]{Nagaosa_book}%
  \BibitemOpen
  \bibfield  {author} {\bibinfo {author} {\bibfnamefont {N.}~\bibnamefont
  {Nagaosa}},\ }\href@noop {} {\emph {\bibinfo {title} {Quantum Field Theory in
  Strongly Correlated Electronic Systems}}}\ (\bibinfo  {publisher} {Springer
  Verlag},\ \bibinfo {year} {1999})\BibitemShut {NoStop}%
\bibitem [{\citenamefont {Ravensbergen}\ \emph {et~al.}(2020)\citenamefont
  {Ravensbergen}, \citenamefont {Soave}, \citenamefont {Corre}, \citenamefont
  {Kreyer}, \citenamefont {Huang}, \citenamefont {Kirilov},\ and\ \citenamefont
  {Grimm}}]{Ravensbergen_2020}%
  \BibitemOpen
  \bibfield  {author} {\bibinfo {author} {\bibfnamefont {C.}~\bibnamefont
  {Ravensbergen}}, \bibinfo {author} {\bibfnamefont {E.}~\bibnamefont {Soave}},
  \bibinfo {author} {\bibfnamefont {V.}~\bibnamefont {Corre}}, \bibinfo
  {author} {\bibfnamefont {M.}~\bibnamefont {Kreyer}}, \bibinfo {author}
  {\bibfnamefont {B.}~\bibnamefont {Huang}}, \bibinfo {author} {\bibfnamefont
  {E.}~\bibnamefont {Kirilov}}, \ and\ \bibinfo {author} {\bibfnamefont
  {R.}~\bibnamefont {Grimm}},\ }\href {\doibase 10.1103/PhysRevLett.124.203402}
  {\bibfield  {journal} {\bibinfo  {journal} {Phys. Rev. Lett.}\ }\textbf
  {\bibinfo {volume} {124}},\ \bibinfo {pages} {203402} (\bibinfo {year}
  {2020})}\BibitemShut {NoStop}%
\bibitem [{\citenamefont {Grest}\ and\ \citenamefont {Sak}(1978)}]{Grest_1978}%
  \BibitemOpen
  \bibfield  {author} {\bibinfo {author} {\bibfnamefont {G.~S.}\ \bibnamefont
  {Grest}}\ and\ \bibinfo {author} {\bibfnamefont {J.}~\bibnamefont {Sak}},\
  }\href {\doibase 10.1103/PhysRevB.17.3607} {\bibfield  {journal} {\bibinfo
  {journal} {Phys. Rev. B}\ }\textbf {\bibinfo {volume} {17}},\ \bibinfo
  {pages} {3607} (\bibinfo {year} {1978})}\BibitemShut {NoStop}%
\bibitem [{\citenamefont {Selke}(1988)}]{Selke_1988}%
  \BibitemOpen
  \bibfield  {author} {\bibinfo {author} {\bibfnamefont {W.}~\bibnamefont
  {Selke}},\ }\href {\doibase https://doi.org/10.1016/0370-1573(88)90140-8}
  {\bibfield  {journal} {\bibinfo  {journal} {Physics Reports}\ }\textbf
  {\bibinfo {volume} {170}},\ \bibinfo {pages} {213 } (\bibinfo {year}
  {1988})}\BibitemShut {NoStop}%
\bibitem [{\citenamefont {Diehl}(2002)}]{Diehl_2002}%
  \BibitemOpen
  \bibfield  {author} {\bibinfo {author} {\bibfnamefont {H.~W.}\ \bibnamefont
  {Diehl}},\ }\href@noop {} {\bibfield  {journal} {\bibinfo  {journal} {Acta
  Physica Slovaca}\ }\textbf {\bibinfo {volume} {52}},\ \bibinfo {pages} {271}
  (\bibinfo {year} {2002})}\BibitemShut {NoStop}%
\bibitem [{\citenamefont {Butera}\ and\ \citenamefont
  {Pernici}(2008)}]{Butera_2008}%
  \BibitemOpen
  \bibfield  {author} {\bibinfo {author} {\bibfnamefont {P.}~\bibnamefont
  {Butera}}\ and\ \bibinfo {author} {\bibfnamefont {M.}~\bibnamefont
  {Pernici}},\ }\href {\doibase 10.1103/PhysRevB.78.054405} {\bibfield
  {journal} {\bibinfo  {journal} {Phys. Rev. B}\ }\textbf {\bibinfo {volume}
  {78}},\ \bibinfo {pages} {054405} (\bibinfo {year} {2008})}\BibitemShut
  {NoStop}%
\bibitem [{\citenamefont {Chaikin}\ and\ \citenamefont
  {Lubensky}(1995)}]{Chaikin_book}%
  \BibitemOpen
  \bibfield  {author} {\bibinfo {author} {\bibfnamefont {P.~M.}\ \bibnamefont
  {Chaikin}}\ and\ \bibinfo {author} {\bibfnamefont {T.~C.}\ \bibnamefont
  {Lubensky}},\ }\href@noop {} {\emph {\bibinfo {title} {Principles of
  condensed matter physics}}}\ (\bibinfo  {publisher} {Cambridge University
  Press},\ \bibinfo {year} {1995})\BibitemShut {NoStop}%
\bibitem [{\citenamefont {Singh}(2000)}]{Singh_2000}%
  \BibitemOpen
  \bibfield  {author} {\bibinfo {author} {\bibfnamefont {S.}~\bibnamefont
  {Singh}},\ }\href {\doibase https://doi.org/10.1016/S0370-1573(99)00049-6}
  {\bibfield  {journal} {\bibinfo  {journal} {Physics Reports}\ }\textbf
  {\bibinfo {volume} {324}},\ \bibinfo {pages} {107} (\bibinfo {year}
  {2000})}\BibitemShut {NoStop}%
\bibitem [{\citenamefont {Goldenfeld}(1992)}]{Goldenfeld_book}%
  \BibitemOpen
  \bibfield  {author} {\bibinfo {author} {\bibfnamefont {N.}~\bibnamefont
  {Goldenfeld}},\ }\href@noop {} {\emph {\bibinfo {title} {Lectures on Phase
  Transitions and the Renormalization Group}}}\ (\bibinfo  {publisher} {Perseus
  Books},\ \bibinfo {year} {1992})\BibitemShut {NoStop}%
\bibitem [{\citenamefont {Pini}\ \emph
  {et~al.}(2021{\natexlab{b}})\citenamefont {Pini}, \citenamefont {Pieri},\
  and\ \citenamefont {Strinati}}]{Pini_2021}%
  \BibitemOpen
  \bibfield  {author} {\bibinfo {author} {\bibfnamefont {M.}~\bibnamefont
  {Pini}}, \bibinfo {author} {\bibfnamefont {P.}~\bibnamefont {Pieri}}, \ and\
  \bibinfo {author} {\bibfnamefont {G.~C.}\ \bibnamefont {Strinati}},\
  }\href@noop {} {\enquote {\bibinfo {title} {Strong fulde-ferrell
  larkin-ovchinnikkov pairing fluctuations in polarized fermi systems},}\ }
  (\bibinfo {year} {2021}{\natexlab{b}}),\ \Eprint
  {http://arxiv.org/abs/2105.00863} {arXiv:2105.00863 [cond-mat.quant-gas]}
  \BibitemShut {NoStop}%
\bibitem [{\citenamefont {Pimenov}\ \emph {et~al.}(2018)\citenamefont
  {Pimenov}, \citenamefont {Mandal}, \citenamefont {Piazza},\ and\
  \citenamefont {Punk}}]{Pimenov_2018}%
  \BibitemOpen
  \bibfield  {author} {\bibinfo {author} {\bibfnamefont {D.}~\bibnamefont
  {Pimenov}}, \bibinfo {author} {\bibfnamefont {I.}~\bibnamefont {Mandal}},
  \bibinfo {author} {\bibfnamefont {F.}~\bibnamefont {Piazza}}, \ and\ \bibinfo
  {author} {\bibfnamefont {M.}~\bibnamefont {Punk}},\ }\href {\doibase
  10.1103/PhysRevB.98.024510} {\bibfield  {journal} {\bibinfo  {journal} {Phys.
  Rev. B}\ }\textbf {\bibinfo {volume} {98}},\ \bibinfo {pages} {024510}
  (\bibinfo {year} {2018})}\BibitemShut {NoStop}%
\bibitem [{\citenamefont {Shpot}\ and\ \citenamefont
  {Diehl}(2001)}]{Shpot_2001}%
  \BibitemOpen
  \bibfield  {author} {\bibinfo {author} {\bibfnamefont {M.}~\bibnamefont
  {Shpot}}\ and\ \bibinfo {author} {\bibfnamefont {H.}~\bibnamefont {Diehl}},\
  }\href {\doibase https://doi.org/10.1016/S0550-3213(01)00309-1} {\bibfield
  {journal} {\bibinfo  {journal} {Nuclear Physics B}\ }\textbf {\bibinfo
  {volume} {612}},\ \bibinfo {pages} {340} (\bibinfo {year}
  {2001})}\BibitemShut {NoStop}%
\bibitem [{\citenamefont {Shpot}\ \emph {et~al.}(2005)\citenamefont {Shpot},
  \citenamefont {Pis'mak},\ and\ \citenamefont {Diehl}}]{Shpot_2005}%
  \BibitemOpen
  \bibfield  {author} {\bibinfo {author} {\bibfnamefont {M.~A.}\ \bibnamefont
  {Shpot}}, \bibinfo {author} {\bibfnamefont {Y.~M.}\ \bibnamefont {Pis'mak}},
  \ and\ \bibinfo {author} {\bibfnamefont {H.~W.}\ \bibnamefont {Diehl}},\
  }\href {\doibase 10.1088/0953-8984/17/20/020} {\bibfield  {journal} {\bibinfo
   {journal} {Journal of Physics: Condensed Matter}\ }\textbf {\bibinfo
  {volume} {17}},\ \bibinfo {pages} {S1947} (\bibinfo {year}
  {2005})}\BibitemShut {NoStop}%
\bibitem [{\citenamefont {Shpot}\ \emph {et~al.}(2008)\citenamefont {Shpot},
  \citenamefont {Diehl},\ and\ \citenamefont {Pismak}}]{Shpot_2008}%
  \BibitemOpen
  \bibfield  {author} {\bibinfo {author} {\bibfnamefont {M.~A.}\ \bibnamefont
  {Shpot}}, \bibinfo {author} {\bibfnamefont {H.~W.}\ \bibnamefont {Diehl}}, \
  and\ \bibinfo {author} {\bibfnamefont {Y.}~\bibnamefont {Pismak}},\ }\href
  {\doibase 10.1088/1751-8113/41/13/135003} {\bibfield  {journal} {\bibinfo
  {journal} {Journal of Physics A: Mathematical and Theoretical}\ }\textbf
  {\bibinfo {volume} {41}},\ \bibinfo {pages} {135003} (\bibinfo {year}
  {2008})}\BibitemShut {NoStop}%
\bibitem [{\citenamefont {Burgsmüller}\ \emph {et~al.}(2010)\citenamefont
  {Burgsmüller}, \citenamefont {Diehl},\ and\ \citenamefont
  {Shpot}}]{Burgsmuller_2010}%
  \BibitemOpen
  \bibfield  {author} {\bibinfo {author} {\bibfnamefont {M.}~\bibnamefont
  {Burgsmüller}}, \bibinfo {author} {\bibfnamefont {H.~W.}\ \bibnamefont
  {Diehl}}, \ and\ \bibinfo {author} {\bibfnamefont {M.~A.}\ \bibnamefont
  {Shpot}},\ }\href {\doibase 10.1088/1742-5468/2010/11/p11020} {\bibfield
  {journal} {\bibinfo  {journal} {Journal of Statistical Mechanics: Theory and
  Experiment}\ }\textbf {\bibinfo {volume} {2010}},\ \bibinfo {pages} {P11020}
  (\bibinfo {year} {2010})}\BibitemShut {NoStop}%
\bibitem [{\citenamefont {Shpot}\ and\ \citenamefont
  {Pismak}(2012)}]{Shpot_2012}%
  \BibitemOpen
  \bibfield  {author} {\bibinfo {author} {\bibfnamefont {M.}~\bibnamefont
  {Shpot}}\ and\ \bibinfo {author} {\bibfnamefont {Y.}~\bibnamefont {Pismak}},\
  }\href {\doibase https://doi.org/10.1016/j.nuclphysb.2012.04.011} {\bibfield
  {journal} {\bibinfo  {journal} {Nuclear Physics B}\ }\textbf {\bibinfo
  {volume} {862}},\ \bibinfo {pages} {75} (\bibinfo {year} {2012})}\BibitemShut
  {NoStop}%
\bibitem [{\citenamefont {{Essafi, K.}}\ \emph {et~al.}(2012)\citenamefont
  {{Essafi, K.}}, \citenamefont {{Kownacki, J.-P.}},\ and\ \citenamefont
  {{Mouhanna, D.}}}]{Essafi_2012}%
  \BibitemOpen
  \bibfield  {author} {\bibinfo {author} {\bibnamefont {{Essafi, K.}}},
  \bibinfo {author} {\bibnamefont {{Kownacki, J.-P.}}}, \ and\ \bibinfo
  {author} {\bibnamefont {{Mouhanna, D.}}},\ }\href {\doibase
  10.1209/0295-5075/98/51002} {\bibfield  {journal} {\bibinfo  {journal} {EPL}\
  }\textbf {\bibinfo {volume} {98}},\ \bibinfo {pages} {51002} (\bibinfo {year}
  {2012})}\BibitemShut {NoStop}%
\bibitem [{\citenamefont {Zappalà}(2017)}]{Zappala_2017}%
  \BibitemOpen
  \bibfield  {author} {\bibinfo {author} {\bibfnamefont {D.}~\bibnamefont
  {Zappalà}},\ }\href {\doibase
  https://doi.org/10.1016/j.physletb.2017.08.051} {\bibfield  {journal}
  {\bibinfo  {journal} {Physics Letters B}\ }\textbf {\bibinfo {volume}
  {773}},\ \bibinfo {pages} {213 } (\bibinfo {year} {2017})}\BibitemShut
  {NoStop}%
\bibitem [{\citenamefont {Zappal\`a}(2018)}]{Zappala_2018}%
  \BibitemOpen
  \bibfield  {author} {\bibinfo {author} {\bibfnamefont {D.}~\bibnamefont
  {Zappal\`a}},\ }\href {\doibase 10.1103/PhysRevD.98.085005} {\bibfield
  {journal} {\bibinfo  {journal} {Phys. Rev. D}\ }\textbf {\bibinfo {volume}
  {98}},\ \bibinfo {pages} {085005} (\bibinfo {year} {2018})}\BibitemShut
  {NoStop}%
\bibitem [{\citenamefont {Defenu}\ \emph {et~al.}(2021)\citenamefont {Defenu},
  \citenamefont {Trombettoni},\ and\ \citenamefont {Zappalà}}]{Defenu_2021}%
  \BibitemOpen
  \bibfield  {author} {\bibinfo {author} {\bibfnamefont {N.}~\bibnamefont
  {Defenu}}, \bibinfo {author} {\bibfnamefont {A.}~\bibnamefont {Trombettoni}},
  \ and\ \bibinfo {author} {\bibfnamefont {D.}~\bibnamefont {Zappalà}},\
  }\href {\doibase https://doi.org/10.1016/j.nuclphysb.2020.115295} {\bibfield
  {journal} {\bibinfo  {journal} {Nuclear Physics B}\ }\textbf {\bibinfo
  {volume} {964}},\ \bibinfo {pages} {115295} (\bibinfo {year}
  {2021})}\BibitemShut {NoStop}%
\bibitem [{\citenamefont {Wetterich}(1993)}]{Wetterich_1993}%
  \BibitemOpen
  \bibfield  {author} {\bibinfo {author} {\bibfnamefont {C.}~\bibnamefont
  {Wetterich}},\ }\href {\doibase https://doi.org/10.1016/0370-2693(93)90726-X}
  {\bibfield  {journal} {\bibinfo  {journal} {Physics Letters B}\ }\textbf
  {\bibinfo {volume} {301}},\ \bibinfo {pages} {90 } (\bibinfo {year}
  {1993})}\BibitemShut {NoStop}%
\bibitem [{\citenamefont {Berges}\ \emph {et~al.}(2002)\citenamefont {Berges},
  \citenamefont {Tetradis},\ and\ \citenamefont {Wetterich}}]{Berges_2002}%
  \BibitemOpen
  \bibfield  {author} {\bibinfo {author} {\bibfnamefont {J.}~\bibnamefont
  {Berges}}, \bibinfo {author} {\bibfnamefont {N.}~\bibnamefont {Tetradis}}, \
  and\ \bibinfo {author} {\bibfnamefont {C.}~\bibnamefont {Wetterich}},\ }\href
  {\doibase https://doi.org/10.1016/S0370-1573(01)00098-9} {\bibfield
  {journal} {\bibinfo  {journal} {Physics Reports}\ }\textbf {\bibinfo {volume}
  {363}},\ \bibinfo {pages} {223 } (\bibinfo {year} {2002})}\BibitemShut
  {NoStop}%
\bibitem [{\citenamefont {Pawlowski}(2007)}]{Pawlowski_2007}%
  \BibitemOpen
  \bibfield  {author} {\bibinfo {author} {\bibfnamefont {J.~M.}\ \bibnamefont
  {Pawlowski}},\ }\href {\doibase https://doi.org/10.1016/j.aop.2007.01.007}
  {\bibfield  {journal} {\bibinfo  {journal} {Annals of Physics}\ }\textbf
  {\bibinfo {volume} {322}},\ \bibinfo {pages} {2831 } (\bibinfo {year}
  {2007})}\BibitemShut {NoStop}%
\bibitem [{\citenamefont {Kopietz}\ \emph {et~al.}(2010)\citenamefont
  {Kopietz}, \citenamefont {Bartosch},\ and\ \citenamefont
  {Sch\"utz}}]{Kopietz_book}%
  \BibitemOpen
  \bibfield  {author} {\bibinfo {author} {\bibfnamefont {P.}~\bibnamefont
  {Kopietz}}, \bibinfo {author} {\bibfnamefont {L.}~\bibnamefont {Bartosch}}, \
  and\ \bibinfo {author} {\bibfnamefont {F.}~\bibnamefont {Sch\"utz}},\
  }\href@noop {} {\emph {\bibinfo {title} {Introduction to the Functional
  Renormalization Group}}}\ (\bibinfo  {publisher} {Springer Verlag},\ \bibinfo
  {year} {2010})\BibitemShut {NoStop}%
\bibitem [{\citenamefont {Polonyi}\ and\ \citenamefont
  {Schwenk}(2012)}]{RG_book}%
  \BibitemOpen
  \bibinfo {editor} {\bibfnamefont {J.}~\bibnamefont {Polonyi}}\ and\ \bibinfo
  {editor} {\bibfnamefont {A.}~\bibnamefont {Schwenk}},\ eds.,\ \href@noop {}
  {\emph {\bibinfo {title} {Renormalization Group and Effective Field Theory
  Approaches to Many-Body Systems}}}\ (\bibinfo  {publisher} {Springer
  Verlag},\ \bibinfo {year} {2012})\BibitemShut {NoStop}%
\bibitem [{\citenamefont {Metzner}\ \emph {et~al.}(2012)\citenamefont
  {Metzner}, \citenamefont {Salmhofer}, \citenamefont {Honerkamp},
  \citenamefont {Meden},\ and\ \citenamefont {Sch\"onhammer}}]{Metzner_2012}%
  \BibitemOpen
  \bibfield  {author} {\bibinfo {author} {\bibfnamefont {W.}~\bibnamefont
  {Metzner}}, \bibinfo {author} {\bibfnamefont {M.}~\bibnamefont {Salmhofer}},
  \bibinfo {author} {\bibfnamefont {C.}~\bibnamefont {Honerkamp}}, \bibinfo
  {author} {\bibfnamefont {V.}~\bibnamefont {Meden}}, \ and\ \bibinfo {author}
  {\bibfnamefont {K.}~\bibnamefont {Sch\"onhammer}},\ }\href {\doibase
  10.1103/RevModPhys.84.299} {\bibfield  {journal} {\bibinfo  {journal} {Rev.
  Mod. Phys.}\ }\textbf {\bibinfo {volume} {84}},\ \bibinfo {pages} {299}
  (\bibinfo {year} {2012})}\BibitemShut {NoStop}%
\bibitem [{\citenamefont {Dupuis}\ \emph {et~al.}(2021)\citenamefont {Dupuis},
  \citenamefont {Canet}, \citenamefont {Eichhorn}, \citenamefont {Metzner},
  \citenamefont {Pawlowski}, \citenamefont {Tissier},\ and\ \citenamefont
  {Wschebor}}]{Dupuis_2021}%
  \BibitemOpen
  \bibfield  {author} {\bibinfo {author} {\bibfnamefont {N.}~\bibnamefont
  {Dupuis}}, \bibinfo {author} {\bibfnamefont {L.}~\bibnamefont {Canet}},
  \bibinfo {author} {\bibfnamefont {A.}~\bibnamefont {Eichhorn}}, \bibinfo
  {author} {\bibfnamefont {W.}~\bibnamefont {Metzner}}, \bibinfo {author}
  {\bibfnamefont {J.}~\bibnamefont {Pawlowski}}, \bibinfo {author}
  {\bibfnamefont {M.}~\bibnamefont {Tissier}}, \ and\ \bibinfo {author}
  {\bibfnamefont {N.}~\bibnamefont {Wschebor}},\ }\href {\doibase
  https://doi.org/10.1016/j.physrep.2021.01.001} {\bibfield  {journal}
  {\bibinfo  {journal} {Physics Reports}\ }\textbf {\bibinfo {volume} {910}},\
  \bibinfo {pages} {1} (\bibinfo {year} {2021})},\ \bibinfo {note} {the
  nonperturbative functional renormalization group and its
  applications}\BibitemShut {NoStop}%
\bibitem [{\citenamefont {De~Polsi}\ \emph {et~al.}(2020)\citenamefont
  {De~Polsi}, \citenamefont {Balog}, \citenamefont {Tissier},\ and\
  \citenamefont {Wschebor}}]{Polsi_2020}%
  \BibitemOpen
  \bibfield  {author} {\bibinfo {author} {\bibfnamefont {G.}~\bibnamefont
  {De~Polsi}}, \bibinfo {author} {\bibfnamefont {I.}~\bibnamefont {Balog}},
  \bibinfo {author} {\bibfnamefont {M.}~\bibnamefont {Tissier}}, \ and\
  \bibinfo {author} {\bibfnamefont {N.}~\bibnamefont {Wschebor}},\ }\href
  {\doibase 10.1103/PhysRevE.101.042113} {\bibfield  {journal} {\bibinfo
  {journal} {Phys. Rev. E}\ }\textbf {\bibinfo {volume} {101}},\ \bibinfo
  {pages} {042113} (\bibinfo {year} {2020})}\BibitemShut {NoStop}%
\bibitem [{\citenamefont {Balog}\ \emph {et~al.}(2019)\citenamefont {Balog},
  \citenamefont {Chat\'e}, \citenamefont {Delamotte}, \citenamefont
  {Marohni\ifmmode~\acute{c}\else \'{c}\fi{}},\ and\ \citenamefont
  {Wschebor}}]{Balog_2019}%
  \BibitemOpen
  \bibfield  {author} {\bibinfo {author} {\bibfnamefont {I.}~\bibnamefont
  {Balog}}, \bibinfo {author} {\bibfnamefont {H.}~\bibnamefont {Chat\'e}},
  \bibinfo {author} {\bibfnamefont {B.}~\bibnamefont {Delamotte}}, \bibinfo
  {author} {\bibfnamefont {M.}~\bibnamefont {Marohni\ifmmode~\acute{c}\else
  \'{c}\fi{}}}, \ and\ \bibinfo {author} {\bibfnamefont {N.}~\bibnamefont
  {Wschebor}},\ }\href {\doibase 10.1103/PhysRevLett.123.240604} {\bibfield
  {journal} {\bibinfo  {journal} {Phys. Rev. Lett.}\ }\textbf {\bibinfo
  {volume} {123}},\ \bibinfo {pages} {240604} (\bibinfo {year}
  {2019})}\BibitemShut {NoStop}%
\bibitem [{\citenamefont {Jakubczyk}\ and\ \citenamefont
  {Wojtkiewicz}(2018)}]{Jakubczyk_2018}%
  \BibitemOpen
  \bibfield  {author} {\bibinfo {author} {\bibfnamefont {P.}~\bibnamefont
  {Jakubczyk}}\ and\ \bibinfo {author} {\bibfnamefont {J.}~\bibnamefont
  {Wojtkiewicz}},\ }\href {\doibase 10.1088/1742-5468/aabc7c} {\bibfield
  {journal} {\bibinfo  {journal} {Journal of Statistical Mechanics: Theory and
  Experiment}\ }\textbf {\bibinfo {volume} {2018}},\ \bibinfo {pages} {053105}
  (\bibinfo {year} {2018})}\BibitemShut {NoStop}%
\bibitem [{\citenamefont {\L{}ebek}\ and\ \citenamefont
  {Jakubczyk}(2020)}]{Lebek_2020}%
  \BibitemOpen
  \bibfield  {author} {\bibinfo {author} {\bibfnamefont {M.}~\bibnamefont
  {\L{}ebek}}\ and\ \bibinfo {author} {\bibfnamefont {P.}~\bibnamefont
  {Jakubczyk}},\ }\href {\doibase 10.1103/PhysRevA.102.013324} {\bibfield
  {journal} {\bibinfo  {journal} {Phys. Rev. A}\ }\textbf {\bibinfo {volume}
  {102}},\ \bibinfo {pages} {013324} (\bibinfo {year} {2020})}\BibitemShut
  {NoStop}%
\bibitem [{\citenamefont {Łebek}\ and\ \citenamefont
  {Jakubczyk}(2021)}]{Lebek_2021}%
  \BibitemOpen
  \bibfield  {author} {\bibinfo {author} {\bibfnamefont {M.}~\bibnamefont
  {Łebek}}\ and\ \bibinfo {author} {\bibfnamefont {P.}~\bibnamefont
  {Jakubczyk}},\ }\href {\doibase 10.21468/SciPostPhysCore.4.2.016} {\bibfield
  {journal} {\bibinfo  {journal} {SciPost Phys. Core}\ }\textbf {\bibinfo
  {volume} {4}},\ \bibinfo {pages} {16} (\bibinfo {year} {2021})}\BibitemShut
  {NoStop}%
\bibitem [{\citenamefont {Chlebicki}\ and\ \citenamefont
  {Jakubczyk}(2021)}]{Chlebicki_2021}%
  \BibitemOpen
  \bibfield  {author} {\bibinfo {author} {\bibfnamefont {A.}~\bibnamefont
  {Chlebicki}}\ and\ \bibinfo {author} {\bibfnamefont {P.}~\bibnamefont
  {Jakubczyk}},\ }\href {\doibase 10.21468/SciPostPhys.10.6.134} {\bibfield
  {journal} {\bibinfo  {journal} {SciPost Phys.}\ }\textbf {\bibinfo {volume}
  {10}},\ \bibinfo {pages} {134} (\bibinfo {year} {2021})}\BibitemShut
  {NoStop}%
\bibitem [{\citenamefont {Canet}\ \emph {et~al.}(2003)\citenamefont {Canet},
  \citenamefont {Delamotte}, \citenamefont {Mouhanna},\ and\ \citenamefont
  {Vidal}}]{Canet_2003_2}%
  \BibitemOpen
  \bibfield  {author} {\bibinfo {author} {\bibfnamefont {L.}~\bibnamefont
  {Canet}}, \bibinfo {author} {\bibfnamefont {B.}~\bibnamefont {Delamotte}},
  \bibinfo {author} {\bibfnamefont {D.}~\bibnamefont {Mouhanna}}, \ and\
  \bibinfo {author} {\bibfnamefont {J.}~\bibnamefont {Vidal}},\ }\href
  {\doibase 10.1103/PhysRevD.67.065004} {\bibfield  {journal} {\bibinfo
  {journal} {Phys. Rev. D}\ }\textbf {\bibinfo {volume} {67}},\ \bibinfo
  {pages} {065004} (\bibinfo {year} {2003})}\BibitemShut {NoStop}%
\bibitem [{\citenamefont {Balog}\ \emph {et~al.}(2020)\citenamefont {Balog},
  \citenamefont {De~Polsi}, \citenamefont {Tissier},\ and\ \citenamefont
  {Wschebor}}]{Balog_2020}%
  \BibitemOpen
  \bibfield  {author} {\bibinfo {author} {\bibfnamefont {I.}~\bibnamefont
  {Balog}}, \bibinfo {author} {\bibfnamefont {G.}~\bibnamefont {De~Polsi}},
  \bibinfo {author} {\bibfnamefont {M.}~\bibnamefont {Tissier}}, \ and\
  \bibinfo {author} {\bibfnamefont {N.}~\bibnamefont {Wschebor}},\ }\href
  {\doibase 10.1103/PhysRevE.101.062146} {\bibfield  {journal} {\bibinfo
  {journal} {Phys. Rev. E}\ }\textbf {\bibinfo {volume} {101}},\ \bibinfo
  {pages} {062146} (\bibinfo {year} {2020})}\BibitemShut {NoStop}%
\end{thebibliography}
\end{document}